\documentclass[prd,showpacs,nofootinbib,preprintnumbers]{revtex4}
\usepackage{graphicx}
\usepackage{amssymb}
\usepackage{pifont}

\def\lQ{\Lambda_{\rm QCD}}
\newcommand{\be}{\begin{equation}}
\newcommand{\ee}{\end{equation}}
\newcommand{\bea}{\begin{eqnarray}}
\newcommand{\eea}{\end{eqnarray}}

\def\siml{{\ \lower-1.2pt\vbox{\hbox{\rlap{$<$}\lower6pt\vbox{\hbox{$\sim$}}}}\ }}

\begin{document}
\title{Non-relativistic bound states in a moving thermal bath}
\author{Miguel Angel Escobedo}
\affiliation{Departament d'Estructura i Constituents de la Mat\`eria and Institut de Ci\`encies del Cosmos, Universitat
de Barcelona\\
Diagonal 647, E-08028 Barcelona, Catalonia, Spain}
\author{Massimo Mannarelli}
\affiliation{Departament d'Estructura i Constituents de la Mat\`eria and Institut de Ci\`encies del Cosmos, Universitat
de Barcelona\\
Diagonal 647, E-08028 Barcelona, Catalonia, Spain}\affiliation{
I.N.F.N., Laboratori Nazionali del Gran Sasso, Assergi (AQ), Italy}
\author{Joan Soto}
\affiliation{Departament d'Estructura i Constituents de la Mat\`eria and Institut de Ci\`encies del Cosmos, Universitat
de Barcelona\\
Diagonal 647, E-08028 Barcelona, Catalonia, Spain}
\preprint{UB-ECM-PF-10/32$\;\;$ ICCUB-11-062}

\begin{abstract}
We study the propagation of non-relativistic bound states moving at constant  velocity across a homogeneous thermal bath 
and we develop the effective field theory which is relevant in various dynamical regimes. We consider values of the 
velocity of the bound state ranging from moderate to highly relativistic and  temperatures at all relevant scales smaller 
than the mass of the particles that form the bound state. In particular, we consider two distinct temperature regimes, 
corresponding to temperatures smaller or higher than the typical momentum transfer in the bound state. For temperatures 
smaller or of the order of the typical momentum transfer, we restrict our analysis to the simplest system, a  
hydrogen-like atom. We build the effective theory for this system first considering  moderate values of the velocity and 
then the relativistic case. For large values of the velocity of the bound state, the separation of scales is such that 
the corresponding effective theory resembles the soft collinear effective theory (SCET). For temperatures larger than the 
typical momentum transfer we also consider muonic hydrogen propagating in a plasma which contains photons and massless
 electrons and positrons, so that the system resembles very much a heavy quarkonium in a thermal medium of deconfined 
quarks and gluons.
We study the behavior of the real and imaginary part of the static two-body potential, for various velocities of the bound
 state, in the hard thermal loop approximation. We find that Landau damping ceases to be the 
relevant mechanism for dissociation from a certain ``critical" velocity on in favor of screening. Our results are 
relevant 
for understanding how the properties of heavy quarkonia 
states produced in the initial fusion of partons in the  relativistic collision of heavy ions are affected by the 
presence of an equilibrated quark-gluon plasma. 
\end{abstract}

\maketitle

\section{Introduction}
When matter is immersed in  a thermal medium  many of its properties change. In principle, no strictly stationary bound 
state exists, because interactions with the particles of the medium lead to a finite lifetime for all states (including 
the ground state). This is equivalent to a broadening  of the  energy levels, {\it i.e.} an imaginary part of the energy 
eigenvalues,  which depends on the density and on the temperature of the medium.

Of particular interest is the case in which the bound state moves with respect to the thermal medium.
The first experimental investigations and theoretical developments of this system  were done in condensed matter 
physics~\cite{cond-mat}.  From the analysis of atoms moving across a plasma,  it has been shown that a number of 
phenomena may take place.   First of all the  Debye screening of the Coulomb potential depends on the relative velocity 
between the bound state and the plasma.  Moreover, the propagation of a bound state through the medium produces a 
 fluctuation of the induced potential which leads to a  density trail. Finally, the moving particle loses energy and is 
eventually stopped by the plasma.

A renewed interest in the properties of bound states moving in a thermal medium arose in recent years due to the advent of
 high energy heavy-ion colliders. In particular, one is interested in understanding whether  some modifications of the 
properties of heavy quarkonia (HQ) states produced in the early stage of the heavy-ion collision can be a signature of  
the presence of a deconfined plasma of quarks and gluons.  
In their pioneering work~\cite{Matsui:1986dk}, Matsui and Satz  showed that the Debye screening of the color interaction 
between two static heavy quarks may lead to  the dissociation of HQ in a thermal medium. This effect should be  
experimentally detectable by the suppression of the corresponding yields. The  suppression of HQ states means that 
 the yield of HQ observed in heavy-ion collisions is smaller than the yield of HQ one  would obtain multiplying HQ 
production rates in p-p collisions by the number of nucleons participating in the collision and taking into account 
the normal nuclear absorption; see {\it e.g.}~\cite{Lourenco:2006sr} for a brief review. The first study of moving HQ 
was then performed in~\cite{Chu:1988wh}, in which the dependence of the  Debye mass  on the velocity of propagation of 
the heavy quarks with respect to the quark-gluon plasma (QGP) was determined.
Subsequent analyses have confirmed the effect and studied the formation  of wakes in the 
QGP~\cite{Abreu:2007kv,  Mustafa:2004hf, Ruppert:2005uz, Chakraborty:2006md, Chakraborty:2007ug, Jiang:2010zza}.

One may wonder whether the drift of bound 
states is important, as it is the case for heavy flavors. Indeed, measurements of heavy flavor production in 
PHENIX~\cite{Awes:2008qi} via single electron measurements results 
in a large $v_2$, which suggest that there is significant damping of heavy quarks while they travel across the fireball. 
Therefore, in heavy-ion collisions the thermal bath expansion may drift the heavy quarks in a phenomenon similar to 
advection in normal fluids. This picture has also received support from 
microscopic calculations of heavy quark diffusion in the quark-gluon plasma~\cite{vanHees:2007me}.
However, we expect that the drag of a heavy quarkonium is less important than that of a heavy quark. This is
 because an isolated heavy quark has a net color charge while a heavy quarkonium at distances larger than its radius is 
colorless. Hence, in general, we expect 
that the HQ states produced in the early times of the collision will not be comoving with the thermal medium, and, 
therefore,
 our calculations will be relevant for them. On the other hand HQ states produced through recombination are expected to 
roughly comove with the thermal
 bath. This is because both heavy quarks have been drifted by the QGP before recombining in a HQ.

Suppression of the J$/\Psi$ was first observed at the CERN SPS~\cite{Abreu:1997jh}. However, in contrast with the naive
 Debye screening scenario, further experimental investigation of the J$/\Psi$ yields at PHENIX~\cite{Awes:2008qi}, 
led to the observation of a strong suppression  at forward rapidity rather than at mid-rapidity.  Recently, there have
 been efforts in studying this problem with the use of non-relativistic effective field theories 
(EFTs)~\cite{Escobedo:2008sy,Brambilla:2008cx,Brambilla:2010vq}. The EFT techniques are very useful for problems that 
have different energy scales, as is the case of HQ in a thermal medium. Using these techniques it has been shown that, at least in perturbation theory, the dissociation of bound states is due to the appearance of an imaginary part 
in the potential~\cite{Laine:2006ns,Escobedo:2008sy,Brambilla:2008cx}. 

In the present  paper we study how a moving thermal bath affects the properties of bound states. One of the points is to 
assess whether the results
 obtained in a static medium are modified  when considering the relative motion between the bound state and the thermal 
medium. We consider 
the simplest systems, hydrogen-like atoms  moving at a constant velocity, {\bf v}, across a homogeneous  thermal medium.  
We study two different cases, the first one corresponds to temperatures smaller or of the order of the typical momentum 
transfer, the second one corresponds to temperatures larger than the typical momentum transfer. We always assume that 
the temperature is much smaller than the mass of the particles forming the bound state.

In the first case we restrict ourselves to the hydrogen atom. We consider separately temperatures 
 smaller than the typical momentum transfer and temperatures of the order of the typical momentum transfer. 
In both temperature ranges we provide the matching procedure and evaluate the  energy shifts and decay widths for the 
stationary states of the system. We build the effective theory for this system first considering  moderate values of 
the velocity and then the relativistic case. We  show that 
for 
large values of the velocity, a new separation of scale occurs and a different EFT must be constructed for dealing with 
bound states. In this case the separation of scale is such that the corresponding EFT resembles some aspects of the soft 
collinear effective theory (SCET)~\cite{Bauer:2000ew}.

In the second case, namely, for temperatures larger than the typical momentum transfer, we also consider muonic hydrogen. 
Since the mass of the particles 
forming the bound state is much larger than the mass of the particles in the thermal bath, this system resembles very much a heavy quarkonium in a thermal medium of deconfined quarks and gluons. As a consequence, this part of the present work is 
of direct relevance for understanding how the properties of heavy quarkonia states produced in the initial fusion of partons in the  
relativistic collision of heavy ions are affected by the presence of an equilibrated quark-gluon plasma. We study the
 behavior of the real and imaginary part of the two-body potential for various values of the  velocity of the bound state 
with respect to the thermal bath employing  the hard thermal loop (HTL) approximation. Regarding the real part of the
 potential we reproduce known results, and extend them to higher speeds. The imaginary part of the potential is calculated 
for the first time. We demonstrate 
that screening overtakes Landau damping as the dominant mechanism for dissociation at a certain critical velocity.

This paper is organized as follows. In Section~\ref{secgen} we introduce some general remarks about the propagation of 
particles in a thermal bath. In Section~\ref{seclowv} and Section~\ref{sechighv} we study the hydrogen atom 
moving at moderate velocities and ultrarelativistic velocities with respect to the medium respectively. 
In Section~\ref{secphoton} we study the real and imaginary part of the static potential of muonic hydrogen in a moving 
thermal bath for temperatures larger than the typical momentum transfer.
The results of this last section are 
directly applicable to the HQ case. Finally, we present our conclusions in Section \ref{secconclusions}.

\section{General framework} \label{secgen}
In our study we shall employ a reference frame in which the bound state is at rest and the thermal medium moves with a 
velocity ${\bf v}$. We choose this frame because it facilitates the application of non-relativistic effective field theories,  in particular, Non-Relativistic
QED~\cite{Caswell:1985ui} (NRQED) and potential NRQED~\cite{Pineda:1997bj} (pNRQED)\footnote{Although the Lagrangian of NRQED is known
for an arbitrary reference frame, most of the developments have been carried out in the rest frame of the bound states.}.
These EFTs are extremely convenient to handle the three different scales of non-relativistic systems at vanishing temperature~\cite{Pineda:1997ie,Pineda:1998kn}, and have already proved useful to analyze these systems in a static thermal bath, in which additional scales occur \cite{Escobedo:2008sy,Escobedo:2010tu}. 
They allow one to organize the calculations in such a way that only one scale 
is taken into account at each step, which, together with the use of dimensional regularization, makes computations much easier.

We shall assume that the plasma (or black-body radiation) is in  thermal equilibrium at a temperature $T$.  Since we are 
considering the reference frame in which the plasma is  moving with a velocity $\bf v$ the particle distribution functions
are given by
\be\label{boosted-distributions}
f (\beta^\mu k_\mu ) =\frac{1}{e^{|\beta^\mu k_\mu|}\pm1},
\ee
where the plus (minus) sign refers to fermions (bosons). In the reference frame where the thermal bath is at rest
$\beta^\mu k_\mu=\frac{k_0}{T}$, while  in a frame where the plasma moves with a velocity $\bf v$ we have that
\be 
\beta^\mu = \frac{\gamma}{T}(1,{\bf v})= \frac{u^\mu}{T} \,, 
\ee
where $\gamma= 1/\sqrt{1-v^2}$, $v=|{\bf v}|$, is the Lorentz factor. 
This frame has been successfully used in the past, for example in~\cite{Weldon:1982aq}.
Studying a bound state in a moving thermal bath is akin to study a bound state in  non-equilibrium field theory~\cite{Carrington:1997sq};
in that case the Bose-Einstein or Fermi-Dirac distribution functions are substituted by a general distribution, 
which in our case 
will be the boosted Bose-Einstein or Fermi-Dirac distribution functions reported in  Eq.~(\ref{boosted-distributions}). 

The vector $\bf v$ brings in the problem a number of complications. First of all, it breaks rotational invariance. Second, when $v$ gets close to $1$
new scales are induced, which has serious implications for the use of EFTs. For instance, if we are working with an EFT in which we have 
integrated out all scales larger than $\mu$, we can no longer argue that if $\mu\gg T$ the Lagrangian of this EFT is not
affected by the temperature. This is because the Boltzmann suppression is not only controlled by $T$, but rather is  a non-trivial function 
of $T$ and $\bf v$. In order to illustrate this point, let us analyze the distribution functions in Eq.~(\ref{boosted-distributions}) in more detail.

We begin with  a thermal bath  consisting of massless particles.
Taking into account that in non-equilibrium field theory the collective behavior always enters through on-shell particles or antiparticles,  we have (in the case of particles) that
\be
\label{dist}
\beta^\mu k_\mu=k\frac{1- v \cos{\theta}}{T\sqrt{1-v^2}}\,,
\ee
where $k=\vert {\bf k}\vert$ and $\theta$ is the angle between $\bf{k}$ and $\bf{v}$. The distribution functions in
Eq.~(\ref{boosted-distributions}) can now be written as 
\be
\label{boosted-2} f(k,T, \theta, v)=\frac{1}{e^{k/T_{\rm eff}(\theta,v)}\pm 1}\,,
\ee
where we have defined the {\it effective temperature}
\be\label{effective-temperature}
T_{\rm eff}(\theta,v)=\frac{T\sqrt{1-v^2}}{1-v\cos{\theta}}\,.
\ee

Intuitively, the dependence of the effective temperature on $v$ and $\theta$ can be understood as a Doppler effect. Indeed Eq.~(\ref{effective-temperature}) is analogous to the change in the frequency of light caused by the relative motion of the source and the observer.
Consider a particle in a thermal bath of radiation moving with velocity $\bf v$. From the point of view of the
particle it will see in the forward direction  blueshifted radiation and in the backward direction redshifted radiation. This corresponds, respectively, to an effective temperature which is higher in the forward direction than in the  backward direction. Therefore,  the effective temperature corresponds to the 
temperature of the radiation as measured by the moving observer and by a minor abuse of language we shall talk about blueshifted and redshifted temperatures.  Notice that analogously to the relativistic Doppler effect, the effective temperature in the transverse direction, {\it i.e.} in the direction corresponding to $\theta=\pi/2$, is redshifted. 

For $v \ll 1$,  one has that  $T_{\rm eff}(\theta,v) \sim T$ for any value of $\theta$ and one single scale $T$ controls the Boltzmann factor in Eq.~(\ref{boosted-2}).
However, for $v$ close to $1$, the values of $T_{\rm eff}(\theta,v)$ strongly depend on $\theta$, which gives rise to an  interesting case for an EFT analysis.  In order to proceed further,
it is convenient to use light-cone coordinates. We choose $\bf v$  in the $z$ direction and define
\be k_+=k_0+k_3 \qquad {\rm and}  \qquad k_-=k_0-k_3 \,.
\label{k+-}
\ee
Then, we have  that
\begin{equation}
\beta^\mu k_\mu=\frac{1}{2}\left(\frac{k_+}{T_+}+\frac{k_-}{T_-}\right) \,,
\label{exp}
\end{equation}
where
\be\label{temperatures}
T_+=T\sqrt{\frac{1+v}{1-v}} \qquad {\rm and} \qquad T_-=T\sqrt{\frac{1-v}{1+v}} \,.
\ee
Therefore, in light cone coordinates, it becomes explicit that the distribution function actually depends on two scales, $T_+$ and $T_-$. For any  value of $v$ it is clear that $T_+ \ge T \ge T_-$ and moreover $T_+$ correspond to the highest temperature measurable by the observer, while $T_-$ corresponds to the lowest temperature measurable by the observer.

For small values of $v$,
$T_+ \simeq T_-$ and the shift in the temperature in the forward and backward directions are negligible. In this case no further separation of scale is needed and one has a single temperature scale,   $T$, as mentioned before.
For $v \simeq 1$, we have $T_+\gg T_-$, namely, two well separated temperature scales, which must be properly taken into account in our EFTs.
Note that configurations with light-cone momenta such that $k_+\gg T_+$ or $k_-\gg T_-$ are exponentially suppressed. Then we can separate the 
remaining configurations in
two regions (in light-cone momenta):
\begin{itemize}
\item A collinear region, corresponding to $k_+\sim T_+$ and $k_-\lesssim T_-$.
\item An ultrasoft region, corresponding to $k_+\ll T_+$ and $k_-\lesssim T_-$.
\end{itemize}
The existence of these two regions has to be taken into account in the matching procedure between different
EFTs. In this paper we shall analyze two different situations in which $v$ is close to $1$, the case $m_e\gg T_+\sim 1/r\gg T_-\gg E$ and the case $T_+\sim m_e\gg 1/r \gg T_-\gg E$.

We would like to remark that although  in Eq.~(\ref{dist})  we have assumed for simplicity that the particles of the 
thermal bath are massless,   our approach applies to  a plasma that consists of   both  massless and massive particles.
Note that our discussion, from Eq. (\ref{k+-}) on, holds independently of what the dispersion 
relation of the particles in the thermal bath is.
If some particles in the plasma have  mass $M\gg T$, we know that 
they are 
exponentially suppressed in the thermal bath, and this must be true in any reference frame. We can easily verify it by 
substituting $k_-=(M^2+{\bf k}^2_{\perp})/k_+$  in Eq.~(\ref{exp}), and by noticing that its minimum is attained at 
 $\beta^\mu k_\mu=M/T$, which confirms that for $M\gg T$ thermal effects due to these particle can indeed be neglected.

\section{Hydrogen atom  at moderate velocities}\label{seclowv}
In the present section  
we shall assume that the velocity is moderate, say $v \lesssim 0.5$, so that $T_+\simeq T \simeq T_-$, and 
separately study the cases $T\ll 1/r$ and $T\sim 1/r$ ($r$ is the size of the bound state, and hence $1/r$ 
of the order of the typical momentum transfer). For simplicity, we also assume that the proton is infinitely heavy.

As we have explained in the previous section, in a thermal bath at a temperature $T\ll M$, particles with mass $M$ are 
exponentially suppressed independently of the value of $v$. In particular, if $M\sim m_e$ indicates the mass  of electrons
 and positrons in the plasma, then these particles are irrelevant in our analysis. In this range of temperature the  hard 
thermal loop  effects  will not appear and the doubling of degrees of freedom plays no role~\cite{Escobedo:2008sy}. 
This implies that only the transverse photons are sensitive to thermal effects and, hence, the leading order interaction,
namely, the Coulomb potential, will not be modified. 

\subsection{The $T\ll 1/r$ case}
\label{Tll1/r}
As a starting point we consider the  pNRQED Lagrangian at vanishing temperature (we use the form given in
Eq. (6) of \cite{Pineda:1997ie}). We will be able to evaluate the corrections to the binding energy $E_n$ and to the decay 
width $\Gamma_n$ due to the 
thermal bath up to the order  $m_e\alpha^5$. There are two different diagrams that contribute at this order. The first one is the  tad-pole diagram which is  given by
\begin{equation}\label{tadpole1}
\parbox{40mm}{\includegraphics[scale=0.3]{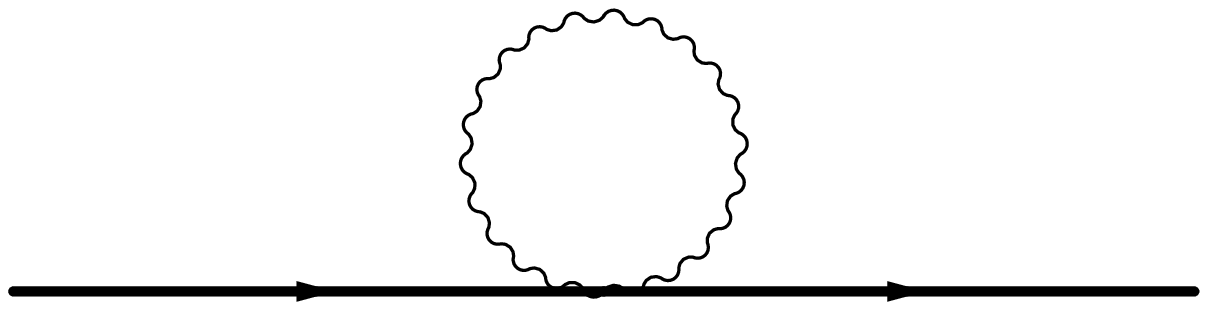}}=-\frac{ie^2}{m_e}
\int\frac{\,d^Dk}{(2\pi)^{D-1}}\frac{\delta(k_0^2-k^2)}{e^{|\beta^\mu k_\mu|}-1}=-\frac{i\pi\alpha T^2}{3m_e}\,,
\end{equation}
where the solid lines represent the atom propagator and the wavy line corresponds to the photon propagator.
In the integral we set the number of  dimensions $D=4$, because the integral is convergent and $k^\mu$ corresponds to  the loop momentum. Notice that the contribution of this diagram is independent of  $\bf v$ 
because the loop integral has no indices and no external momentum enters in it. 
This is in fact true for any tad-pole diagram of this kind. Therefore one can read the result from the ${\bf v}=0$ case 
\cite{Escobedo:2008sy}.

Next we consider the ``rainbow" diagram, 
\begin{equation}\label{rainbow1}
\parbox{40mm}{\includegraphics[scale=0.6]{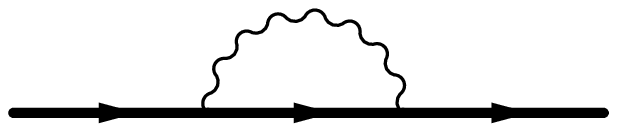}}=-\frac{e^2}{m_e^2}\lim_{p_0\to E_n}\sum_r\langle n|p^i|r\rangle I_{ij}(p_0-E_r)\langle r|p^j|n\rangle\,.
\end{equation}
$p^i$ is the momentum operator of the electron, $|r\rangle$ symbolizes an eigenstate of the Coulomb Hamiltonian 
of energy $E_r$, and $I_{ij}(q)$ is defined as follows:
\begin{equation}\label{Iijq}
I_{ij}(q)=\int\frac{\,d^Dk}{(2\pi)^{D-1}}\frac{\delta(k_0^2-k^2)}{e^{|\beta^\mu k_\mu|}-1}\frac{i}{q-k_0+i\epsilon}\left(\delta_{ij}-\frac{k_ik_j}{k^2}\right)\,. 
\end{equation}

Since the only two independent tensors  are $\delta_{ij}$ and $v_iv_j$, one can use the decomposition 
\begin{equation}\label{Iij}
I_{ij}=A\,P^s_{ij}+B\,P^p_{ij} \,,
\end{equation}
where 
\begin{equation}\label{projectors}
P^s_{ij}=\frac{1}{2}\left(\delta_{ij}+\frac{v_iv_j}{v^2}\right) \qquad {\rm and} \qquad P^p_{ij}=\frac{1}{2}\left(\delta_{ij}-3\frac{v_iv_j}{v^2}\right) \,,
\end{equation}
and then 
\begin{equation}
A=\frac{1}{2}I_{ii} \qquad {\rm and} \qquad  B=\frac{1}{2}\left(I_{ii}-2\frac{v^iv^j}{v^2}I_{ij}\right) \,.
\end{equation}
Upon substituting  Eq.~(\ref{Iijq}) in the expressions above,  we find that 
\begin{equation}
\label{relA}
A=\int\frac{\,d^Dk}{(2\pi)^D}\frac{\delta(k_0^2-k^2)}{e^{|\beta^\mu k_\mu|}-1}\frac{i}{q-k_0+i\epsilon} \qquad {\rm and}\qquad B=\int\frac{\,d^Dk}{(2\pi)^D}\frac{\delta(k_0^2-k^2)}{e^{|\beta^\mu k_\mu|}-1}\frac{i}{q-k_0+i\epsilon}\frac{({\bf v}\cdot{\bf k})^2}{v^2 k^2}\,. 
\end{equation}
The computations of these integrals is done in Appendix \ref{apen1}. The results for the imaginary and real part of $A$ are, respectively, given by
\begin{equation}\label{ImA}
\Im A(q)=\frac{q}{4\pi^2}\left(\log\left(\frac{2\pi T}{|q|}\right)+1-\frac{1}{2v}\log\left(\frac{1+v}{1-v}\right)+\frac{\pi T\sqrt{1-v^2}}{|q|v}\Im\log\left(\frac{\Gamma\left(\frac{i|q|}{2\pi T_+}\right)}{\Gamma\left(\frac{i|q|}{2\pi T_-}\right)}\right)\right)
\end{equation}
and
\begin{equation}\label{ReA}
\Re A(q)=-\frac{T\sqrt{1-v^2}}{8\pi v}\log\left(\frac{1-e^{-\frac{|q|}{T_+}}}{1-e^{-\frac{|q|}{T_-}}}\right) \,.
\end{equation}
The imaginary part of $B$ turns out to be  given by
\begin{equation}\label{ImB}
\Im B(q)=\frac{q}{12\pi^2}\left(\log\left(\frac{2\pi T}{|q|}\right)+\frac{1}{3v^2}(3+v^2)-\frac{1}{2v^3}\log\left(\frac{1+v}{1-v}\right)+\frac{3}{2}\int_{-1}^1\,d\lambda\lambda^2\Re\Psi\left(\frac{i|q|(1-v\lambda)}{2\pi T\sqrt{1-v^2}}\right)\right) \,,
\end{equation}
while for the real part of $B$ we have
\begin{equation}\label{ReB}
\Re B(q)=-\frac{T\sqrt{1-v^2}}{8\pi v^3}\int_{\frac{|q|}{T_-}}^{\frac{|q|}{T_+}}\,dt\left(1-\frac{T\sqrt{1-v^2}t}{|q|}\right)^2\frac{1}{e^t-1} \,.
\end{equation}
The expressions above can be computed numerically for any value of the parameters.
The thermal corrections to the energy and the decay width (for arbitrary angular momentum) 
are given by the following expressions:
\begin{equation}
\delta E_{nlm}=\frac{\alpha \pi T^2}{3m}+\frac{e^2}{m_e^2}\lim_{p_0\to E_n}\sum_r\langle n|p^i|r\rangle\Im I_{ij}(p_0-E_r)\langle r|p^j|n\rangle\,,
\label{E_IIIA}
\end{equation}
and
\begin{equation}
\delta\Gamma_{nlm}=\frac{2e^2}{m_e^2}\lim_{p_0\to E_n}\sum_r\langle n|p^i|r\rangle\Re I_{ij}(p_0-E_r)\langle r|p^j|n\rangle\,,
\label{G_IIIA}
\end{equation}
which in general will depend on the relative velocity $\bf v$. Analytical expressions for  $T\gg E$ and for  $T\ll E$, 
where $E\sim E_n$, the binding energy scale, are derived below. 

\subsubsection{The  $T\gg E$ case}
For  $T\gg E$  the leading contribution to the  integrals in Eqs.~(\ref{ImA}), (\ref{ReA}) and Eqs.~(\ref{ImB}), (\ref{ReB}) can be analytically determined  and upon substituting the corresponding expressions in  $I_{ij}$  we find that
\be
\label{rei}
\Re I_{ij}=\frac{T\sqrt{1-v^2}}{8\pi v}\left(P^s_{ij}\log\left(\frac{1+v}{1-v}\right)+P^p_{ij}\frac{\log\left(\frac{1+v}{1-v}\right)-2v}{v^2}\right)+\mathcal{O}(E) \,,
\ee
for the real part of $I_{ij}$ and
\be
\label{imi}
\Im I_{ij}=\frac{q}{4\pi^2}\left[P^s_{ij}\left(\log\frac{2\pi T}{|q|}-\frac{1}{2v}\log\left(\frac{1+v}{1-v}\right)+1-\gamma\right)+P^p_{ij}\frac{1}{3}\left(\log\frac{2\pi T}{|q|}+\frac{1}{v^2}+\frac{1}{3}-\frac{1}{2v^3}\log\left(\frac{1+v}{1-v}\right)-\gamma\right)\right]+\mathcal{O}\left(\frac{E^2}{T}\right) \,,
\ee
for the imaginary part of $I_{ij}$.

It is interesting to observe that, in this limit, the terms non-local in $q$ (Bethe-log type) coincide with those obtained for vanishing velocity. Hence, all  dependence on $\bf v$ is encoded in an anisotropic potential and kinetic term. 
This result will also serve as a cross-check of
the $1/r\sim T$ computation 
that will be carried out in the next section.

In order to obtain the energy shift and the decay width from Eqs.~(\ref{rei}) and (\ref{imi}) we consider separately 
the S-wave states and states with non-vanishing angular momentum. We display below slightly more general results, 
which hold for an ion of charge $Z$ as well.

In the S-wave states the expected value of any tensor operator $\langle n|O_{ij}|n\rangle\propto\delta_{ij}$, therefore  terms proportional to $P^p_{ij}$, can be ignored (because  $P^p_{ij}$ is traceless) and we find that
\begin{eqnarray}
\delta E_{n}&=&\frac{\alpha\pi T^2}{3m_e}-\frac{4Z\alpha^2}{3}\frac{|\phi_n({\bf 0})|^2}{m_e^2}\left(-\frac{1}{2v}\log\left(\frac{1+v}{1-v}\right)+1-\gamma\right) \nonumber\\
&+&\frac{2\alpha}{3\pi m_e^2}\sum_r|\langle n|{\bf p}|r\rangle|^2(E_n-E_r)\log\left(\frac{2\pi T}{|E_n-E_r|}\right), 
\label{E0}
\end{eqnarray} 
where $\phi_n({\bf 0})$ is the wave function at the origin. The corresponding change in the width turns out to be given by 
\begin{equation}
\delta\Gamma_{n}=\frac{2Z^2\alpha^3T\sqrt{1-v^2}}{3n^2 v}\log\left(\frac{1+v}{1-v}\right)\,.
\label{G0}
\end{equation}

States with non-vanishing angular momentum are more difficult to deal with. It is convenient to decompose $I_{ij}$ by the tensors $\delta_{ij}$ and $\frac{v_iv_j}{v^2}$ instead of $P^s_{ij}$ and $P^p_{ij}$. For the real and imaginary parts of $I_{ij}$ we find respectively 
\begin{equation}
\Re I_{ij}=\frac{T\sqrt{1-v^2}}{16\pi v}\left[\delta_{ij}\left(\left(1+\frac{1}{v^2}\right)\log\left(\frac{1+v}{1-v}\right)-\frac{2}{v}\right)+\frac{v_iv_j}{v^2}\left(\left(1-\frac{3}{v^2}\right)\log\left(\frac{1+v}{1-v}\right)+\frac{6}{v}\right)\right],
\end{equation}
and
\begin{eqnarray}
\Im I_{ij}&=&\frac{q}{4\pi}\left[\frac{2\delta_{ij}}{3}\left(\log\left(\frac{2\pi T}{|q|}\right)-\gamma+\frac{5}{6}+\frac{1}{4v^2}-\frac{3}{8v}\left(1+\frac{1}{3v^2}\right)\log\left(\frac{1+v}{1-v}\right)\right)-\frac{v_iv_j}{2v^2} \rho(v)\right]\,,
\end{eqnarray}
where
\begin{equation}\label{rhov}
\rho(v)=\frac{1}{2v}\left(1-\frac{1}{v^2}\right)\log\left(\frac{1+v}{1-v}\right)-\frac{2}{3}+\frac{1}{v^2} \,.
\end{equation}
To determine the energy shifts and the decay widths we fix ${\bf v}$ in the $z$-direction, and make use of the following identities 
\begin{equation}
\Bigg< n\Bigg|\frac{({\bf v}\cdot{\bf p})^2}{v^2}\Bigg|n\Bigg>=\langle n|p^2|n\rangle\frac{\sqrt{4\pi}}{3}\Bigg< Y_{lm}\Bigg|\left(Y_{00}+\sqrt{\frac{4}{5}}Y_{20}\right)\Bigg|Y_{lm}\Bigg>\,,
\end{equation}
and
\begin{equation}
\langle n|[[H,{\bf v}\cdot {\bf p}],{\bf v}\cdot{\bf p}]|n\rangle=\sqrt{\frac{4\pi}{5}}Z\alpha\Bigg< n\Bigg|\frac{1}{r^3}\Bigg|n\Bigg>\langle Y_{lm}|Y_{20}|Y_{lm}\rangle,
\end{equation}
where $Y_{lm}$ are the spherical harmonics. Note that $|n\rangle$ is used as a short-hand notation for $|nlm\rangle$, where $n$ is the principal quantum number, $l$ the orbital angular momentum and $m$ its  $z$-component.  With these expressions we obtain the general forms of the shifts of the energy levels
\begin{eqnarray}
\delta E_{nlm}&=&\frac{\alpha \pi T^2}{3m_e}+\frac{2\alpha}{3\pi m_e^2}\sum_r|\langle n|{\bf p}|r\rangle|^2(E_n-E_r)\log\left(\frac{-E_1}{|E_n-E_r|}\right) \nonumber\\
&-&\frac{Z^3\alpha^2\langle 2l00|l0\rangle\langle 2l0m|lm\rangle}{2\pi m_e^2a_0^3n^3l(l+\frac{1}{2})(l+1)}\rho(v)\,, 
\label{El}
\end{eqnarray}
and the corresponding  widths are given by
\begin{equation}
\delta\Gamma_{nlm}=\frac{Z^2\alpha^3T\sqrt{1-v^2}}{3n^2v}\left(2\log\left(\frac{1+v}{1-v}\right)+\left(\left(1-\frac{3}{v^2}\right)\log\left(\frac{1+v}{1-v}\right)+\frac{6}{v}\right)\langle 2l00|l0\rangle\langle 2l0m|lm\rangle\right)\,,
\label{Gl}
\end{equation}
where $\langle lml'm'|l''m''\rangle$ are the Glebsch-Gordan coefficients 
(normalization and sign conventions are as in~\cite{Sakurai}).

It is interesting to observe that the decay widths (\ref{G0}) and (\ref{Gl}) decrease as the velocity increases.

\subsubsection{The  $ T \ll E$ case}
For temperatures  $ T\ll E$  the coefficients $A$ and $B$  simplify and upon replacing their expressions  in Eq.~(\ref{Iij})  we obtain that \begin{equation}
I_{ij}(q)=\frac{iT^2(1-v^2)}{32q}\int_{-1}^1\frac{\,d\lambda}{(1-v\lambda)^2}\left(P_{ij}^s+\lambda^2P_{ij}^p\right)+\mathcal{O}\left(\frac{T^4}{q^3}\right) \,.
\end{equation}
When this expression is used in the
evaluation of the energy shift in Eq.~(\ref{E_IIIA})  and of the decay width in
Eq.~(\ref{G_IIIA}), we need to calculate
\begin{equation}
 P_{ij}^x\sum_r\langle n|p^i|r\rangle\frac{1}{E_n-E_r}\langle
 r|p^j|n\rangle=-{m_e\over 2} P_{ii}^x \,,
\end{equation}
for $x=s,p$. Note that only the term proportional to $P_{ij}^s$ contributes because $P_{ij}^p$ is traceless.
 Thus, the contribution from the rainbow diagram in the limit $T\ll E$ for states with vanishing angular momentum is given by
\begin{equation}
\parbox{40mm}{\includegraphics[scale=0.3]{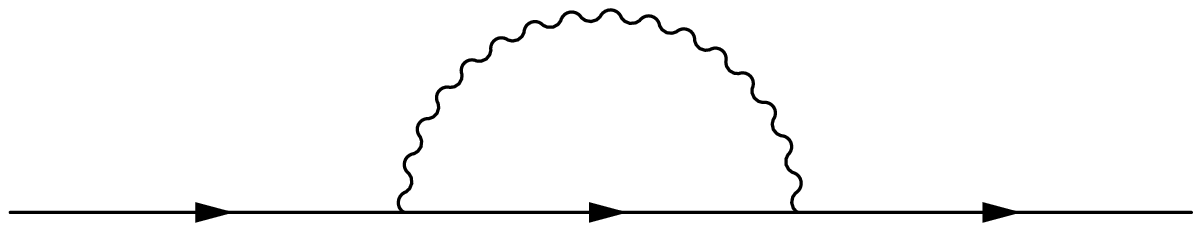}}=\frac{i\pi\alpha T^2}{3m_e}+\mathcal{O}\left(\frac{\alpha^3T^4}{E^3}\right)\,,
\end{equation}
and is independent of the velocity $v$. Hence, the dominant contribution of the 
rainbow diagram above cancels the contribution of the tad-pole diagram, which we have seen 
to be independent of $v$ as well. Then the thermal corrections  in this case are very 
suppressed, of the order  ${\cal O}(\alpha^3T^4/E^3)$, like in the case of the thermal bath at rest.

\subsection{The $T\sim 1/r$ case}

In  the temperature regime $T\sim 1/r$, the temperature is high enough so that its effects  must be taken into account already 
in the matching between NRQED and pNRQED, namely it affects the potential. Therefore, several diagrams are modified by 
the presence of the temperature.
However, like in the calculations for the thermal bath at rest,
there are only four diagrams that give a relevant contribution. All other diagrams give contributions that either vanish 
or can be  shown to cancel out by local field redefinitions. We schematically analyze the relevant diagrams below.
Then, as a cross check, 
we compare the calculations for $T \sim 1/r$ in the limit of low temperature, with the results that  we derived in the 
previous section for $E\sim T$ in the limit of high temperature, and we find agreement. 

The four diagrams that must be taken into account in the matching procedure between NRQED and pNRQED are the following:

\begin{itemize}
\item The tad-pole diagram which comes from the $\frac{D^2}{2m_e}$ term in the NRQED Lagrangian and gives the contribution
\begin{equation}
\parbox{40mm}{\includegraphics[scale=0.3]{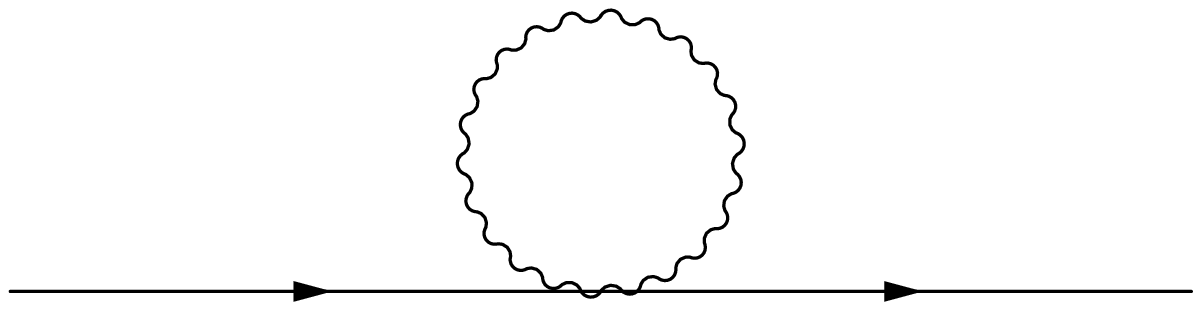}}=-\frac{i\pi\alpha T^2}{3m_e}\,.
\end{equation}
This diagram is quite similar to  the diagram we evaluated in the previous section, but now the solid line corresponds to the electron field instead of the hydrogen atom. As already discussed, this diagram is independent of the velocity $v$.

\item The rainbow diagram is given by
\begin{eqnarray}
\parbox{40mm}{\includegraphics[scale=0.3]{./selfus.eps}}&=&-\frac{-ie^2p^i p^j}{2m_e^2}\int\frac{\,d^{D-1}k}{(2\pi)^{D-1}k}\frac{1}{e^{\frac{k}{T\sqrt{1-v^2}}(1-\frac{{\bf v}\cdot{\bf k}}{k})}-1}\left(\delta_{ij}-\frac{k_ik_j}{k^2}\right) \\
&\times&\left(\frac{1}{p_0-k-\frac{({\bf p}-{\bf k})^2}{2m_e}}+\frac{1}{p_0+k-\frac{({\bf p}+{\bf k})^2}{2m_e}}\right) \nonumber \,,
\end{eqnarray}
where the solid line corresponds to the electron field and the wavy line corresponds to the photon.
In order to have a consistent  EFT we have to expand
\be
\label{expansion}
\frac{1}{p_0-k-\frac{({\bf p}-{\bf k})^2}{2m_e}}+\frac{1}{p_0+k-\frac{({\bf p}+{\bf k})^2}{2m_e}}\to -2\frac{p_0-\frac{p^2}{2m_e}}{k^2}-\frac{1}{m_e}\,,
\ee
and therefore the contribution of the rainbow diagram can be written as
\begin{equation}
\label{qedtensor}
\parbox{40mm}{\includegraphics[scale=0.3]{./selfus.eps}}=\left(p_0-\frac{p^2}{2m_e}\right) \frac{ie^2p^ip^j}{m_e^2}T_{ij}+\frac{ie^2}{2m_e^3}p^ip^jR_{ij}\,,
\end{equation}
where
\begin{equation}\label{Tij}
T_{ij}=\int\frac{\,d^{D-1}k}{(2\pi)^{D-1}k^3}\frac{1}{e^{\frac{k}{T\sqrt{1-v^2}}(1-\frac{{ \bf v}\cdot{\bf k}}{k})}-1}\left(\delta_{ij}-\frac{k_ik_j}{k^2}\right)\,,
\end{equation}
and
\begin{equation}\label{Rij}
R_{ij}=\int\frac{\,d^{D-1}k}{(2\pi)^{D-1}k}\frac{1}{e^{\frac{k}{T\sqrt{1-v^2}}(1-\frac{{ \bf v}\cdot{\bf k}}{k})}-1}\left(\delta_{ij}-\frac{k_ik_j}{k^2}\right)\,.
\end{equation}
For the time being we do not evaluate these integrals;  as we shall clarify soon we only need  to evaluate $T_{ij}$. 
\item The  thermal correction of the Coulomb potential  corresponds to the  diagram  
\begin{equation}
\label{rdcou}
\parbox{40mm}{\includegraphics[scale=0.3]{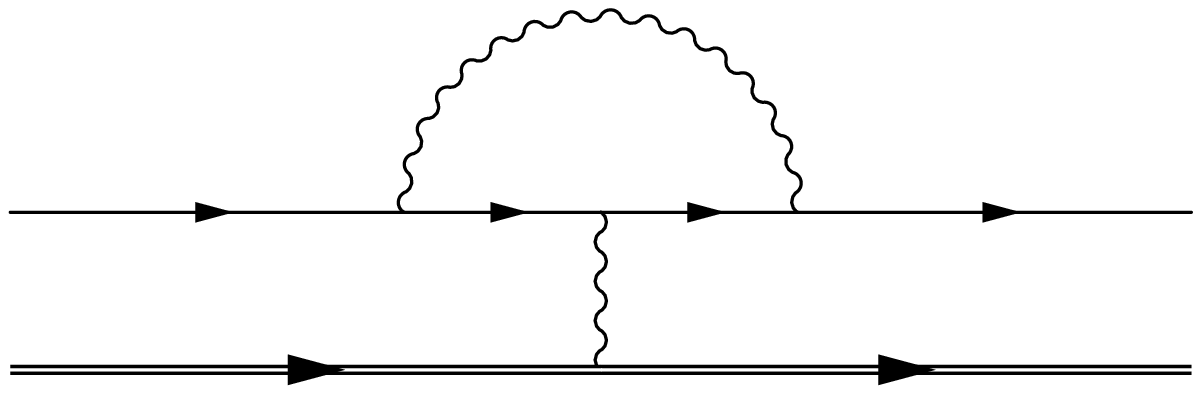}}=\frac{ie^4p^i{p'}^j}{m_e^2|{\bf p}-{\bf p}'|^2}T_{ij}\,,
\end{equation}
where the tensor $T_{ij}$ is the same as in Eq. (\ref{Tij}) and $\bf p$ and $\bf p^\prime$ are the momenta of the incoming and outgoing electrons, respectively.
The solid thick line here corresponds to the ion propagator and the solid thin line is the electron propagator.

\item The last diagram to consider is the relativistic tad-pole, that is the same as the previous tad-pole diagram, but now the vertex comes from the $\frac{D^4}{8m_e^3}$ term in the NRQED Lagrangian. The contribution of this diagram is given by
\begin{equation}
\parbox{40mm}{\includegraphics[scale=0.3]{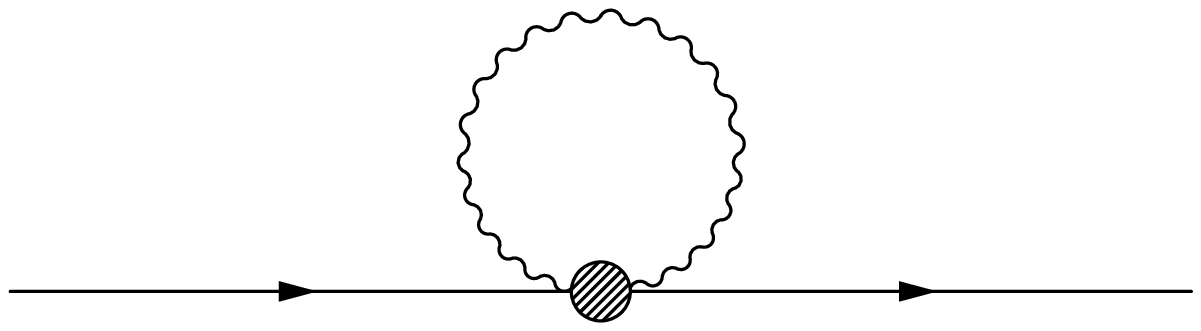}}=\frac{ip^2\pi\alpha T^2}{6m_e^3}-\frac{ie^2}{2m_e^3}p^ip^jR_{ij}\,,
\end{equation}
where the tensor $R_{ij}$ is defined in Eq.~(\ref{Rij}). Notice that the term on the right hand side of Eq.~(\ref{qedtensor}) cancels the corresponding contribution of this diagram. Therefore, the sum of the rainbow diagram and of the relativistic  tad-pole diagram is independent of $R_{ij}$. 
\item The remaining non-vanishing diagrams give contributions analogous to the ones  of Eqs.~(36), (37) and (42) of~\cite{Escobedo:2008sy}. 
Their net effect can be shown to be zero by local field redefinitions, like in the case of the thermal bath at rest.
\end{itemize}
Now, let us consider how these diagrams combine. In particular, we would like to obtain  the pNRQED Lagrangian that matches all these terms. By inspection we  find that the pNRQED Lagrangian is given by
\begin{eqnarray}
\delta \mathcal{L}_{pNRQED}&=&\int\,d^3{\bf x}\left(\frac{4\pi\alpha}{m_e^2}T_{ij}\left[\frac{\partial^2_{ik}\psi^\dagger\partial^2_{jk}\psi}{2m_e}+\partial^2_{ij}\psi^\dagger\partial_0\psi\right] + \frac{\alpha \pi T^2}{3m_e}\psi^\dagger\psi - \frac{\pi\alpha T^2}{6m_e^3}{\bf\nabla}\psi^\dagger{\bf \nabla}\psi\right) \nonumber \\
&+&\int\,d^3{\bf x_1}\,d^3{\bf x_2}N^\dagger(t,{\bf x_2}) N(t,{\bf x_2})\frac{4\alpha \pi}{m_e^2}\frac{Z\alpha}{|{\bf x_1}-{\bf x_2}|}T_{ij}\nabla^i\psi^\dagger(t,{\bf x_1})\nabla^j\psi(t,{\bf x_1})\,,
\label{L0}
\end{eqnarray} 
which as already noted does not depend on $R_{ij}$. 
We evaluate $T_{ij}$ in a similar way as was done for $I_{ij}$ in the $T\sim E$ case. The result in the $\overline{MS}$ 
subtraction scheme is
\be
T_{ij}=\frac{1}{6\pi^2}\left[\log\left(\frac{\mu}{T}\right)-\log 2-\log(2\pi)+\gamma+\frac{3}{8v}\left(1+\frac{1}{3v^2}\right)\log\left(\frac{1+v}{1-v}\right)-\frac{1}{4v^2}\right]\delta_{ij} + \frac{\rho(v)}{8\pi^2}  \frac{v_iv_j}{v^2} \,,
\label{tij}
\ee
where  we have decomposed  $T_{ij}$ in terms of $\delta_{ij}$ and $v_iv_j/v^2$, and $\rho(v)$ is defined in Eq.~(\ref{rhov}). 
Upon making a local field redefinition to remove the term with a time derivative in (\ref{L0}),
one can identify the corrections to the potential, in a similar way as it was done in Eq.~(45) of~\cite{Escobedo:2008sy}. The final form of the thermal correction to the pNRQED Lagrangian reads
\begin{eqnarray}
\delta\mathcal{L}_{pNRQED}&=&\int\,d^3{\bf x}\left(\frac{\alpha \pi T^2}{3m_e}\psi^\dagger\psi-\frac{\pi\alpha T^2}{6m_e^3}{\bf\nabla}\psi^\dagger{\bf \nabla}\psi\right) 
+\int\,d^3{\bf x_1}\,d^3{\bf x_2}N^\dagger(t,{\bf x_2}) N(t,{\bf x_2})\left[-\frac{4Z\alpha}{3m_e^2}\left(\log\left(\frac{\mu}{2\pi T}\right)-\log 2 \right.\right.\nonumber\\ &+&\left.\left. \gamma + \frac{3}{8v}\left(1+\frac{1}{3v^2}\right)\log\left(\frac{1+v}{1-v}\right)-\frac{1}{4v^2}\right)\delta^3({\bf x_1}-{\bf x_2})
+\frac{\alpha\rho(v)v^iv^j\partial^2_{ij}V_c(r)}{4\pi m_e^2v^2}\right]\psi^\dagger(t,{\bf x_1})\psi(t,{\bf x_1})\,,
\end{eqnarray}
where $V_c(r)$ is the Coulomb potential at vanishing temperature.
With simple modifications this Lagrangian can be put in a form so that we have an atom field instead of an electron and a nucleus field (see~\cite{Escobedo:2008sy} for more details).

Now that we have computed the corrections to the pNRQED Lagrangian for the case $T\sim 1/r$ we also need to compute the contribution from the ultrasoft scale with this Lagrangian. These contributions can be computed  from the tad-pole and the rainbow  diagrams as in Eqs.(\ref{tadpole1}) and (\ref{rainbow1}), where  the Bose-Einstein distribution function can be expanded because we are now in the case $T\gg E$, and therefore
\begin{equation}
\frac{1}{e^{|\beta^\mu k_\mu|}-1 } \to \frac{1}{|\beta^\mu k_\mu|}-\frac{1}{2}+\cdots \,.
\end{equation}
Upon substituting the expansion above in Eq.~(\ref{tadpole1}) we find that the contribution of the tadpole diagram vanishes in dimensional regularization, because it has no scales (it is independent of the external momentum).
The rainbow diagram gives a contribution similar to the one in Eq.~(\ref{rainbow1}), with the replacement  $I_{ij} \to J_{ij}$, where
\be
\Re J_{ij}=\frac{T\sqrt{1-v^2}}{8\pi v}\left(P^s_{ij}\log\left(\frac{1+v}{1-v}\right)+P^p_{ij}\frac{\log\left(\frac{1+v}{1-v}\right)-2v}{v^2}\right) \,,
\label{ReJij}
\ee
and
\begin{equation}\label{ImJij}
\Im J_{ij}(q)=\frac{q}{6\pi^2}\delta_{ij}\left(\log\left(\frac{\mu}{|q|}\right)+\frac{5}{6}-\log 2\right)\,.
\end{equation}
In order to be consistent with (\ref{tij}), the $\overline{MS}$ scheme has also been used here to 
remove the UV divergences.
The thermal corrections to the energy levels and to the decay width  coming from the soft and the ultrasoft scales in the case $T\sim 1/r$ are respectively given by 
\begin{eqnarray}
\delta E_{nlm} &=&\frac{\alpha\pi T^2}{3m_e}-\frac{\pi\alpha^3T^2}{6m_en^2}+\frac{4Z\alpha^2}{3m_e^2}\left(\log\left(\frac{\mu}{4\pi T}\right)+\gamma+\frac{3}{8v}\left(1+\frac{1}{3v^2}\right)-\frac{1}{4v^2}\right)|\phi_n({\bf 0})|^2 \nonumber\\
&-&\frac{\alpha\rho(v)v^iv^j}{4\pi m_e^2v^2}\langle n|\partial^2_{ij}V_c(r)|n\rangle+\frac{e^2}{m_e^2}\lim_{p_0\to E_r}\sum_r\langle n|p^i|r\rangle \Im J_{ij}(p_0-E_r)\langle r|p^j|n\rangle\,,
\label{E_IIIB}
\end{eqnarray}
and
\begin{equation}
\delta\Gamma_{nlm}=\frac{2e^2}{m_e^2}\lim_{p_0\to E_n}\sum_r\langle n|p^i|r\rangle\Re J_{ij}(p_0-E_r)\langle r|p^j|n\rangle\,.
\label{G_IIIB}
\end{equation}
Note that the $\mu$ dependence in the correction to the Darwin term in (\ref{E_IIIB}) is canceled out by the $\mu$ 
dependency of $J_{ij}(q)$ in Eq.~(\ref{ImJij}).  
Note also that, upon substituting (\ref{ReJij}) in (\ref{G_IIIB}), the expression for the decay width
reduces to that of (\ref{Gl}). Furthermore,
the thermal corrections to the 
binding energy above  
in the limit of low temperature coincide with (\ref{E0}) and (\ref{El}).
 Therefore,  the limit of low temperature 
in the case $T\sim 1/r$ agrees with the limit of high temperature in the $T\sim E$ case.

\section{Hydrogen atom at relativistic velocities} \label{sechighv}
When the bound state moves at high speed with respect to  the thermal bath one has to take into account that the effective
 temperature measured by the bound state in the forward direction is  blueshifted and that the effective temperature  
in the backward direction  is redshifted.  In particular one has that $T_+\gg T_-$, and therefore  $T_+$ and $T_-$, 
which we have defined in  Eq.~(\ref{temperatures}), are two well separated energy scales that must be properly taken 
into account in the  EFT.  In particular it is possible that $T_+$ and $T_-$ are in two distinct energy ranges. In this 
case  the analysis of the system differs considerably with respect to  the case of the thermal bath at rest. 
We shall  study two  different 
situations of this sort: the first one corresponds to the case $T_+\sim 1/r\gg T_-\gg E$ and the second one to the case   $T_+\sim m_e\gg 1/r\gg T_-\gg E$. 
Recall that we are assuming that $T \ll m_e$ and hence, as we have already stressed in Section~\ref{secgen}, we can neglect positrons and electrons in the thermal bath.

\subsection{The $T_+\sim 1/r\gg T_-\gg E$ case}
Since $ T_+ \ll m_e$ we can use  NRQED at vanishing temperature as the starting point, but we would like to integrate 
out also the $1/r$ scale in order to construct the  pNRQED for this situation. As in the soft collinear effective theory, 
it is convenient to split the photon field $A_\mu$ into two different components, a collinear one $A_\mu^{\rm col}$ that 
takes into account photons with $k_+\sim T_+$ and $k_-\sim T_-$ and a ultrasoft one $A_\mu^{\rm us}$ that takes into
 account photons with $k_+\sim k_-\sim T_-$ (or smaller). Notice that both types of photons have virtualities $\lambda$ 
that fulfill $(1/r)^2\gg \lambda$; this means that neither of these two types of photons have to be integrated out in 
the matching between NRQED and pNRQED. We carry out the matching below using an electron field and a nucleus field in pNRQED 
(rather than an atom field), as it was done 
in \cite{Pineda:1997ie}.

\subsubsection{Matching between NRQED and pNRQED for collinear photons}
\label{mcol}
\begin{figure}
\includegraphics[scale=0.5]{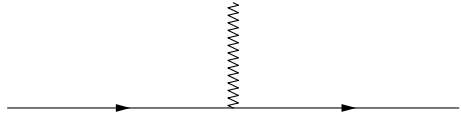}
\caption{Minimal coupling  in pNRQED between an electron and a collinear photon. The solid lines represent the electron propagators, the zigzag line represents the collinear photon. This diagram is not kinematically allowed because when  a non-relativistic electron interacts with an ultrasoft photon its virtuality changes by a quantity of the order $m_e/r$. Then, the outgoing electron cannot be described within pNRQED. }
\label{fignopos}
\end{figure}

In the matching procedure between NRQED and pNRQED we have to determine the effective vertex between non-relativistic
 electrons and collinear photons.  In pNRQED the interaction of non-relativistic electrons with collinear photons cannot
 be given by the minimal coupling diagram shown in Fig.~\ref{fignopos}
for kinematical reasons, as we argue next. Let us call $p$ the momentum of the incoming electron and $k$ the momentum 
of the incoming collinear photon. The non-relativistic electron in pNRQED 
must have $p_0-\frac{p^2}{2m_e}\ll 1/r$, because $m_e/r$ is precisely the typical virtuality of the electrons that 
have been integrated out in the matching between NRQED and pNRQED. However, if the photon is collinear, namely
 $k_0 \sim 1/r$, then  
the virtuality of the outgoing electron is of order $m_e/r$, in contradiction with the fact that electrons with 
such a virtuality do not appear in pNRQED.

\begin{figure}
\includegraphics[scale=0.5]{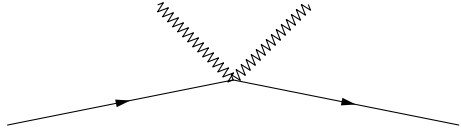}
\caption{Leading-order term for the interaction of electrons with collinear photons in pNRQED.}
\label{figpos}
\end{figure}

Therefore, the interaction between electrons and collinear photons in pNRQED is given at leading order by 4-point processes as the one presented in Fig.~\ref{figpos}. The matching procedure is outlined in the following equation
\begin{equation}
\label{match1}
\parbox{40mm}{\includegraphics[scale=0.30]{./4point.eps}}+\frac{1}{2}\left(\parbox{40mm}{\includegraphics[scale=0.3]{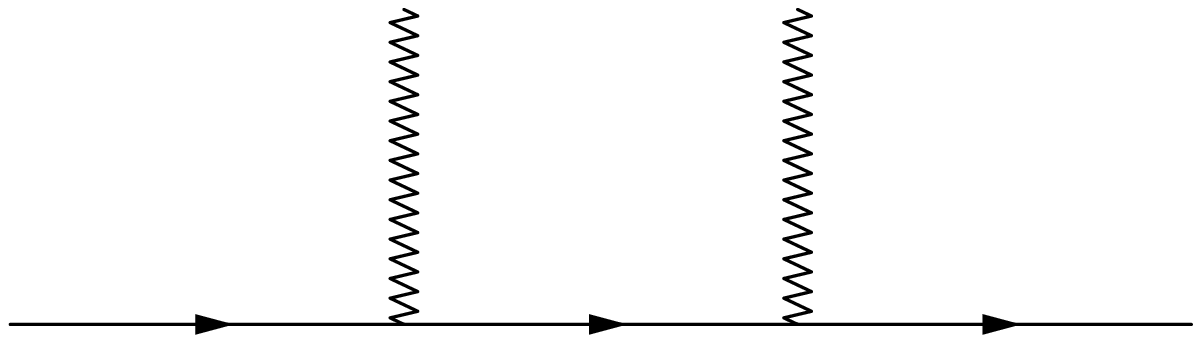}}+\parbox{40mm}{\includegraphics[scale=0.3]{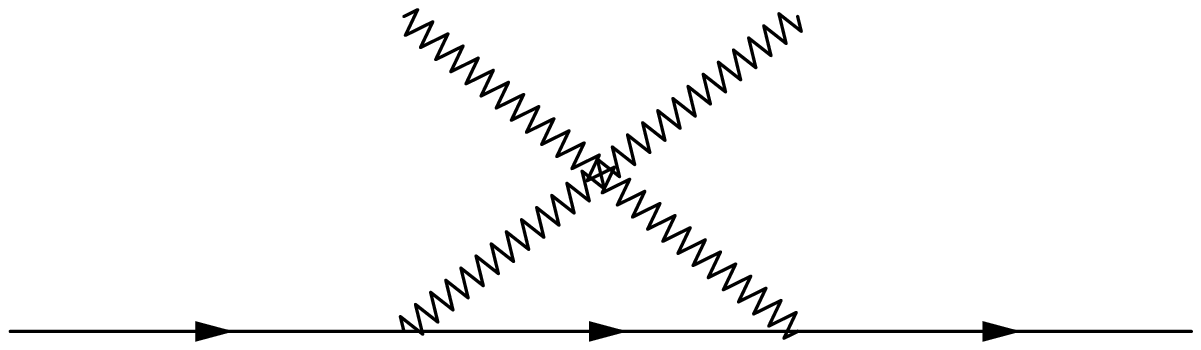}}\right)=\parbox{40mm}{\includegraphics[scale=0.3]{./4point.eps}}
\end{equation}
The NRQED diagrams on the left hand side have to match the pNRQED diagram on the right hand side. 
This part of 
the pNRQED Lagrangian involving collinear photons is given at the required order in Appendix \ref{ape:pnrqed}.  
The part of the pNRQED Lagrangian  
involving ultrasoft photons only is the same as in the case with the thermal bath at rest.

\subsubsection{Computation using pNRQED}
Since we have determined the pNRQED Lagrangian, we can now calculate the contribution of thermal collinear photons to the self-energy of the hydrogen atom [from the part of the Lagrangian reported in Eq. (\ref{lagrascet1})]. We shall use  the Coulomb gauge and therefore only spatial components contribute. Moreover, the condition $\bf{\nabla}\cdot \bf{A}=0$ for collinear photons means that $A_3(x)\ll A_\perp(x)$ and   we only need the first and second terms of Eq.~(\ref{lagrascet1}). The contribution of collinear photons  to the self-energy of the hydrogen atom is given by
\begin{equation}
\parbox{40mm}{\includegraphics[scale=0.3]{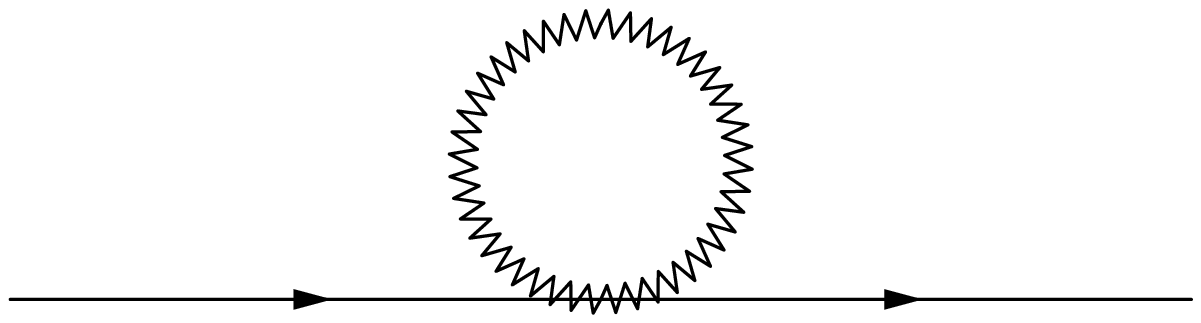}}=-\frac{ie^2}{m_e}\int\frac{\,d^Dk}{(2\pi)^{D-1}}\frac{\delta(k_+k_--k_\perp^2)}{e^{\frac{1}{2}|\frac{k_+}{T_+}+\frac{k_-}{T_-}|}-1}=-\frac{i\pi\alpha T^2}{3m_e}\,,
\end{equation}
where the zig-zag line corresponds to a collinear photon and the solid line represents the hydrogen atom. 
As already noticed in the discussion of Eq.~(\ref{tadpole1}), the tad-pole diagram is not sensitive to the relative motion between the bound state and the thermal bath, because no external momenta enter into the loop. The corresponding  shift to the  energy levels  is given by
\begin{equation}
\delta E^{\rm col}=\frac{\pi\alpha T^2}{3m_e}\,.
\label{TEc}
\end{equation}
As we shall see soon, the contribution coming from thermal collinear photons is the dominant one, however it does not depend on the quantum numbers of the state and therefore it cannot be seen in the emission spectra.

There are two one-loop contributions of ultrasoft photons to the self-energy of the hydrogen atom. The tad-pole contribution   is given by 
\begin{equation}
\parbox{40mm}{\includegraphics[scale=0.3]{./tadus.eps}}=-\frac{ie^2}{m_e}\int\frac{\,d^Dk}{(2\pi)^{D-1}}\frac{\delta(k_+k_--k_\perp^2)}{e^{\frac{|k_-|}{2T_-}}-1}=-\frac{ie^2}{m_e}\frac{\Omega_{D-2}}{(2\pi)^{D-1}}\int_{-\infty}^\infty\frac{\,dk_-}{|k_-|(e^{\frac{|k_-|}{2T_-}}-1)}\int_0^\infty\,dk_\perp k_\perp^{D-3}\,,
\end{equation}
and we find that in dimensional regularization this integral vanishes. The second contribution is due to the rainbow diagram
\begin{equation}
\label{uscon}
\parbox{40mm}{\includegraphics[scale=0.3]{./selfus.eps}}=-\frac{e^2}{m_e^2}\lim_{p_0\to E_n}\sum_r\langle n|p^i|r\rangle K_{ij}(p_0-E_r)\langle r|p^j|n\rangle\,,
\end{equation}
where
\begin{equation}
K_{ij}=\int\frac{\,d^Dk}{(2\pi)^{D-1}}\frac{\delta(k_+k_--k_\perp^2)}{e^{\frac{1}{2}|\frac{k_+}{T_+}+\frac{k_-}{T_-}|}-1}\left(\delta_{ij}-\frac{k_ik_j}{k^2}\right)\frac{i}{q-\frac{1}{2}(k_++k_-)+i\epsilon}\,.
\end{equation}
There are three different integration regions that contribute to this integral 
\begin{itemize}
\item the region with $k_+,k_-\sim T_-$\,,
\item the region with $k_+,k_-\sim q$\,, and
\item the region with $k_+\sim q$ and $k_-\sim q(T_-/T_+)$\,.
\end{itemize}
Note that $k_+/T_+\ll 1$ in all the regions. Evaluating the contributions of each region (see Appendix \ref{apen2}) 
and putting them together we find that 
\be\label{defK}
K_{ij}=a P^s_{ij}+b P^p_{ij} \,,
\ee
where $P^s_{ij}$ and $ P^p_{ij}$ are defined in Eq.~(\ref{projectors}) and 
\begin{eqnarray}
\Re a &=& \frac{T}{4\pi}\sqrt{\frac{1-v}{1+v}}\log\left(\frac{1+v}{1-v}\right)\,, \\
\Im a &=&\frac{q}{4\pi^2}\left(1-\gamma+\log\left(\frac{2\pi T}{q}\right)+\frac{1}{2}\log\left(\frac{1-v}{1+v}\right)\right)\,,
\end{eqnarray}
and 
\begin{eqnarray}
\Re b &=& \frac{T}{4\pi}\sqrt{\frac{1-v}{1+v}}\left(\log\left(\frac{1+v}{1-v}\right)-2\right)\,, \\
\Im b &=& \frac{q}{12\pi^2}\left(\frac{4}{3}-\gamma+\log\left(\frac{2\pi T}{q}\right)+\frac{1}{2}\log\left(\frac{1-v}{1+v}\right)\right)\,.
\end{eqnarray}
Note that in the $v\to 1$ limit,  the  coefficients $a$ and $b$ are equal to  the coefficients $A$ and $B$ reported, respectively, in Eqs.(\ref{ImA}), (\ref{ReA}) and  Eqs.(\ref{ImB}), (\ref{ReB}) in the limit $q\ll T_-$. Therefore, in the same limit, 
we have that $K_{ij} \to I_{ij}$. The thermal corrections to the energy and decay widths due to the ultrasoft photons can be written as
\begin{equation}
\delta E_{nlm}^{\rm us}=\frac{e^2}{m_e^2}\lim_{p_0\to E_n}\sum_r\langle n|p^i|r\rangle \Im K_{ij}(p_0-E_r)\langle r|p^j|n\rangle\,,
\end{equation}
and
\begin{equation}
\delta \Gamma_{nlm}^{\rm us}=\frac{2e^2}{m_e^2}\lim_{p_0\to E_n}\sum_r\langle n|p^i|r\rangle \Re K_{ij}(p_0-E_r)\langle r|p^j|n\rangle\,.
\end{equation}
For S-wave states we obtain the following  expressions for the energy shifts: 
\begin{equation}
\delta E_{n0}^{\rm us}=-\frac{4Z\alpha^2}{3}\left(1-\gamma+\frac{1}{2}\log\left(\frac{1-v}{1+v}\right)\right)\frac{|\phi_n(0)|^2}{m_e^2}-\frac{2\alpha}{3\pi m_e^2}\sum_r|\langle n|{\bf p}|r\rangle|^2(E_n-E_r)\log\left(\frac{|E_n-E_r|}{2\pi T}\right),
\label{TEus0}
\end{equation}
and decay widths
\begin{equation}
\delta\Gamma_{n0}^{\rm us}=\frac{4Z^2\alpha^3 T}{3n^2}\sqrt{\frac{1-v}{1+v}}\log\left(\frac{1+v}{1-v}\right).
\label{Twus0}
\end{equation}
For states with non-vanishing angular momentum $l$ we find that
\begin{equation}
\delta E_{nlm}^{\rm us}=
\frac{2\alpha}{3\pi m_e^2}\sum_r|\langle n|{\bf p}|r\rangle|^2(E_n-E_r)\log\left(\frac{-E_1}{|E_n-E_r|}\right)-\frac{Z^3\alpha^2\langle 2l00|l0\rangle\langle 2l0m|lm\rangle}{6\pi m_e^2a_0^3l(l+\frac{1}{2})(l+1)},
\label{TEusl}
\end{equation}
and
\begin{equation}
\delta\Gamma_{nlm}^{\rm us}=\frac{4Z^2\alpha^3 T}{3n^2}\sqrt{\frac{1-v}{1+v}}\left[\log\left(\frac{1+v}{1-v}\right)-\left(\log\left(\frac{1+v}{1-v}\right)-3\right)\langle 2l00|l0\rangle\langle 2l0m|lm\rangle\right].
\label{Twusl}
\end{equation}
The total thermal width is given by the ultrasoft contribution reported in Eq.~(\ref{Twus0}), for S-wave states, or 
in Eq.~(\ref{Twusl}) for states with non-vanishing angular momentum. In order to obtain the total thermal energy shift 
the collinear contribution given in Eq.~(\ref{TEc}) must be added to the ultrasoft contributions given in 
Eq.~(\ref{TEus0}) for S-wave states, or in Eq.~(\ref{TEusl}) for states with non-vanishing angular momentum. Note that the 
latter turns out to be totally independent of the velocity.

Note that the decay widths (\ref{Twus0}) and (\ref{Twusl}) are decreasing functions of the velocity, like in the 
moderate velocity case. Furthermore, the results above agree with those of Sec. \ref{Tll1/r} in the $v\to 1$ limit. 

\subsection{The $T_+\sim m_e\gg 1/r\gg T_-\gg E$ case}
We shall now consider a highly relativistic hydrogen atom immersed in a thermal bath at a temperature  $T\sim 1/r$. 
We shall assume that the relative velocity between the hydrogen atom and the thermal bath is such that the temperature 
in the forward direction is blueshifted to the electron mass, that is $T_+\sim m_e$, while in the backward direction  
the effective temperature  is redshifted to $1/r \gg T_- \gg E$.  The effective temperatures $T_+$ and $T_-$ are now 
very well separated scales, therefore this situation is specially suitable for the use of EFT. 

In the construction of the effective theory we start with QED at vanishing temperature,  because $m_e\gg T$. However, the existence of collinear photons must be taken into account in the matching between QED and NRQED. On this aspect the matching procedure is akin to the one in SCET. We shall schematically describe this matching procedure below. Regarding collinear photons, they have a virtuality  of order $(1/r)^2$, and they must be integrated out when matching from NRQED to pNRQED.
Finally, the contributions of ultrasoft photons are calculated in pNRQED. In this case pNRQED does not include collinear photons, which already have been integrated out. The interaction with ultrasoft photons is exactly the same as in the previous case. Their contribution is given by exactly the same diagram as in Eq.~(\ref{uscon}) and therefore one obtains the energy shifts reported in Eq.~(\ref{TEus0}), for S-wave states, and in Eq.~(\ref{TEusl}), for states with non-vanishing angular momentum, and the widths are given by the same expressions reported in Eq.~(\ref{Twus0}), for S-wave states, and in Eq.~(\ref{Twusl}), for states with non-vanishing angular momentum.
\subsubsection{Matching between QED and NRQED  for collinear photons}
In QED, when a non-relativistic electron absorbs a collinear photon, it turns into a relativistic electron. This means that the NRQED Lagrangian cannot have this kind of 3-body interaction (a similar argument was used in Section \ref{mcol}). Hence,  in NRQED the interaction with non-relativistic electrons has to be a 4-body interaction.

In this case, there is the additional complication that on the QED side we have bispinors, while in NRQED we have only spinors. This can be solved using the non-relativistic projector. The matching equation takes the form
\begin{equation}
\label{match2}
\frac{1}{2}\frac{1+\gamma_0}{2}\left(\parbox{40mm}{\includegraphics[scale=0.3]{./4point2.eps}}+\parbox{40mm}{\includegraphics[scale=0.3]{./4point3.eps}}\right)\frac{1+\gamma_0}{2}=\sqrt{Z}\frac{1+\gamma_0}{2}\parbox{40mm}{\includegraphics[scale=0.3]{./4point.eps}}\sqrt{Z},
\end{equation}
where $Z$ is the wave function renormalization of NRQED that depends quadratically of the momentum. The result for this matching is given in Appendix \ref{ape:nrqed}.

To match NRQED with pNRQED we have to integrate out collinear photons. The contribution of collinear photons  to the self-energy is given by
\begin{equation}
\parbox{40mm}{\includegraphics[scale=0.3]{./tadcol.eps}}=-\frac{i\pi\alpha T^2}{3m_e}\left(1-\frac{p^2}{2m_e^2}\right)\,,
\end{equation}
where  we have used the Lagrangian of Eq.~(\ref{lagrascet2}) in the Coulomb gauge. Note that in this gauge thermal effects are only due to the spatial components of $A_\mu$ and $A_3\ll A_\perp$, and  we only need the terms proportional to $c_1$, $c_2$, $c_{11}$, and $c_{13}$ reported in Appendix  \ref{ape:nrqed}.
This diagram was already evaluated in the case of the thermal bath at rest. Since tad-pole diagrams are unaffected by the motion of the thermal bath, the result remains the same, and the only effect is  a constant energy shift in the pNRQED Lagrangian, which amounts to the following shift of the effective mass of the electron:
\begin{equation}
\delta m_e = \frac{\pi\alpha T^2}{3m_e}\,.
\label{eshift}
\end{equation}
Regarding heavy quarks, the net effect of collinear gluons is a very tiny shift of the heavy quark mass by an amount of $\delta m_Q \sim \alpha_s T^2/m_Q$, which is irrelevant for the stability analysis of heavy quarkonia. Much more important for the stability analysis is  how the  Coulomb potential changes at high temperatures and this will be studied  in the following section.

\section{The static potential of muonic hydrogen in the range $ T\gg 1/r$}
\label{secphoton}
In  muonic hydrogen the proton  is orbited  by a muon and the bound state consists of two heavy  particles. Since the muon is about 207 times heavier than the electron, muonic hydrogen is much more compact than standard hydrogen and the energy levels of the system have about 207 times the energy of standard hydrogen.   Muonic hydrogen is  investigated in order to have high precision measurements of the proton properties~\cite{review-H}, mainly by Lamb shift measurements~\cite{Pohl:2010}.  The study of muonic atoms is also important for  muon-catalyzed fusion processes~\cite{fusion}, which are under  experimental investigation at RIKEN~\cite{riken} and Star Scientific~\cite{star}.

The study of muonic hydrogen in a thermal bath with  $m_\mu\gg T\gg m_e$ is akin to the study of HQ states in the
quark-gluon plasma
 with $m_Q \gg T\gg \lQ \gg m_q$, where $m_Q$ is the mass of the heavy quark and 
$m_q$ ($q=u,d,s$) is the mass of light quarks \cite{Escobedo:2010tu}. The reason is that in both cases the temperature 
is on the one hand 
much smaller than the masses of the particles that form the bound state and on the other hand much larger 
than the  mass of the particles in the thermal bath. There are thermally excited electrons and positrons in the QED plasma
 and thermally excited light quarks in the QGP which can modify the Coulomb interaction between the two heavy particles. 
Actually,  we shall assume that the temperature is high enough so that we can neglect the masses of the light particles 
of the plasma.

Apart from the modification of the static Coulomb potential,   the propagation of a particle in the medium produces a  fluctuation of the induced potential which leads to a  variation in the density of the plasma. If the plasma behaves as a liquid the moving bound state can produce a wake.  These effects were first analyzed in  condensed matter physics (see {\it e.g.}~\cite{cond-mat}) and then studied in the context of heavy-ion collisions~\cite{Abreu:2007kv, Chu:1988wh, Mustafa:2004hf, Ruppert:2005uz, Chakraborty:2006md, Chakraborty:2007ug, Jiang:2010zza} and in strongly coupled $N=4$ supersymmetric Yang-Mills plasmas~\cite{Chernicoff:2006hi,Peeters:2006iu, Liu:2006nn, Chesler:2007an}.

In the present section we study the modifications to the leading-order potential between two heavy sources in relative 
motion with respect to the thermal bath  at a velocity $\bf v$. We evaluate the potential in the HTL approximation 
assuming that the temperature of the plasma is $T\gg 1/r$. The real part of the potential is screened by massless 
particles loops, both in QED and QCD (the only difference between QED and QCD in our results is, apart from trivial 
color factors, the value of the Debye mass $m_D$).  The real part of the potential between a quark and an antiquark
 moving in a thermal bath was first computed  in the HTL approximation in~\cite{Chu:1988wh} and then more recently 
in~\cite{Chakraborty:2006md}.  Recent perturbative calculations~\cite{Laine:2006ns} at vanishing velocity have pointed 
out the importance of the imaginary part of the potential. So far its effect has not been taken into account 
in a moving thermal bath.

In the Coulomb gauge the potential is obtained by the Fourier transform of the longitudinal photon propagator,  
\be\label{delta11}
\Delta_{11}(k)=\frac{1}{2}[\Delta_R(k)+\Delta_A(k)+\Delta_S(k)] \,,
\ee
for $k_0 \ll \vert {\bf k}\vert $, where $\Delta_R(k)$ and $\Delta_A(k)$ are respectively the retarded and the advanced propagators and  $\Delta_S(k)$ is the symmetric propagator.
For a bound state comoving with the thermal bath, it is enough to compute the retarded self-energy in the rest frame of the thermal bath and then using
\be
\Delta^*_R(k)=\Delta_A(k)\,,
\ee
and 
\be
\Delta_S(k)=[1+2f(|k_0|,T)]{\rm sgn}(k_0)[\Delta_R(k)-\Delta_A(k)]\,,
\ee
one can determine the potential. In the expression above, $f(k_0,T)$  is the distribution function of the longitudinal photons in the thermal bath.
However, the last relation does not hold for a bound state moving through a thermal bath~\cite{Carrington:1997sq}, and must be substituted by the following one:
\be
\label{noeq}
\Delta_S(k,u)=\frac{\Pi_S(k,u)}{2i\Im\Pi_R(k,u)}(\Delta_R(k,u)-\Delta_A(k,u)) \,,
\ee
where $u^\mu=\gamma(1,{\bf v})$ is the 4-velocity. Thus, in order to determine the propagator one has to evaluate the self-energies $\Pi_R(k,u)$ and $\Pi_S(k,u)$.

The retarded self-energy $\Pi_R(k,u)$ was computed in~\cite{Chu:1988wh} and
 here we only show the result in the reference frame where the bound state is at 
rest\footnote{In \cite{Chu:1988wh} there is a misprint in the first line of Eq.~(8), in which the global sign must be the 
opposite.}
\begin{equation}
\Pi_R(k,u)=a(z)+\frac{b(z)}{1-v^2},
\end{equation}
where $z=\frac{v\cos\theta}{\sqrt{1-v^2\sin^2\theta}}$, $\theta$ is the angle between ${\bf k}$ and ${\bf v}$, and
\begin{equation}
a(z)=\frac{m_D^2}{2}\left(z^2-(z^2-1)\frac{z}{2}\ln\left(\frac{z+1+i\epsilon}{z-1+i\epsilon}\right)\right),
\end{equation}
\begin{equation}
b(z)=(z^2-1)\left(a(z)-m_D^2(1-z^2)\left(1-\frac{z}{2}\ln\left(\frac{z+1+i\epsilon}{z-1+i\epsilon}\right)\right)\right).
\end{equation}

Regarding the symmetric self-energy of the longitudinal photons $\Pi_S(k,u)$, the computation is similar to the one 
done for the retarded self-energy in~\cite{Chu:1988wh}. Consider the full symmetric self-energy tensor $\Pi^s_{\mu\nu}$. 
It obeys the Ward  identity 
\be
k^\mu\Pi^s_{\mu\nu}=0\,,
\ee
and is symmetric,
\be\Pi^s_{\mu\nu}=\Pi^s_{\nu\mu} \,.
\ee
Then, it must have the following structure:
\begin{equation}
\Pi^{s\;{\mu\nu}}=\Pi_1\left(g^{\mu\nu}-\frac{k^\mu k^\nu}{k^2 }\right)+\Pi_2 u^\mu_{\perp} u^\nu_{\perp} \,,
\ee
where  $\Pi_1$ and $\Pi_2$ are two scalars and
\be\label{u-perp}
u^\mu_{\perp} = \left(u^\mu-\frac{(k\cdot u)k^\mu}{k^2}\right)\,,
\end{equation}
is the component of $u^\mu$ orthogonal to $k_\mu$. Since $\Pi^s_{\mu \nu}$ is a tensor,  we can determine the values of $\Pi_1$ and $\Pi_2$ in any reference frame, and it is convenient to consider the comoving frame, {\it i.e.} the frame in which the thermal bath is at rest. It is useful to  define the tensor
\begin{equation}
P^{\mu\nu}=\frac{1}{2}\left(u^\mu u^\nu-g^{\mu\nu}+\frac{k^\mu_\perp k^\nu_\perp}{k^2-(k \cdot u)^2}\right) \,,
\end{equation}
where
\be\label{k-perp}
k^\mu_\perp = k^\mu-(k \cdot u)u^\mu 
\ee
is the component of $k^\mu$ orthogonal to $u_\mu$, and 
$k_{\perp}^2= k^2-(k \cdot u)^2$. It is clear that  $P^{\mu \nu}u_\nu = P^{\mu \nu} k_\nu =0$, and therefore $P^{\mu \nu}$ projects four-vectors in the direction orthogonal to  $u^\mu$ and $k^\mu$. By means of Eqs.~(\ref{u-perp}) and (\ref{k-perp}) it is easy to show that $P_{\mu \nu}u^\nu_\perp = P_{\mu \nu} k^\nu_\perp =0$, as well. Then we have that
\begin{equation}
P^{\mu\nu}\Pi^s_{\mu\nu}= P^{\mu\nu} g_{\mu \nu} \Pi_1  = -\Pi_1 \,,
\end{equation}
and in the comoving frame one has that  the only nonvanishing components of $P_{\mu\nu}$ are
\be
P^{ij}=\frac{1}{2}\left(\delta^{ij}-\frac{k^ik^j}{k^2}\right)\,,
\ee
and then  in this frame
\begin{equation}
P^{\mu\nu}\Pi^s_{\mu\nu}=\frac{1}{2}\left(\delta^{ij}-\frac{k^ik^j}{k^2}\right)\Pi^s_{ij}\,,
\end{equation}
which is precisely the transverse component of the photon (gluon) self-energy $\Pi^S_T$ in the Coulomb gauge. This quantity has been computed for vanishing velocity in \cite{Carrington:1997sq} and we find that 
\begin{equation}
\Pi_1=-i\pi m_D^2\frac{T}{\sqrt{(k\cdot u)^2-k^2}}\left(1-\frac{(k \cdot u)^2}{(k \cdot u)^2-k^2}\right)\theta(-k^2)\,.
\end{equation}
The  scalar quantity $u^\mu u^\nu\Pi^s_{\mu\nu}$   has a simple interpretation in the comoving frame, where it turns out to be given by  $\Pi^s_{00}$. Then we have that
\begin{equation}
\Pi_2=\left(\frac{(k\cdot u)^2}{(k\cdot u)^2-k^2}-1\right)\left(\Pi_1-\left(1-\frac{(k\cdot u)^2}{(k\cdot u)^2-k^2}\right)2i\pi m_D^2T\frac{\theta(-k^2)}{\sqrt{(k\cdot u)^2-k^2}}\right)\,,
\end{equation}
and we can now compute the symmetric self-energy in the frame where the muonic hydrogen is at rest and the thermal bath is moving with a velocity $v$. In this frame we have that 
\begin{equation}
\Pi_S(k,u)=\Pi^s_{00}=\Pi_1+\frac{\Pi_2}{1-v^2}=\frac{i2\pi m_D^2 T(1-v^2)^{3/2}(1+\frac{v^2}{2}\cos^2\theta)}{|{\bf k}|(1-v^2\sin^2\theta)^{5/2}}\,.
\label{Pis00}
\end{equation}
We have all the necessary quantities to construct $\Delta_S$ using Eq.~(\ref{noeq}) and then the propagator in 
Eq.~(\ref{delta11}). When the limit $v\to 0$ is taken, we obtain for $\Delta_S$ the same result as in refs. 
\cite{Laine:2006ns,Escobedo:2008sy,Brambilla:2008cx}.

In the previous discussion we have not distinguished between the case of moderate velocities and the case of relativistic velocities, 
as we did in the hydrogen atom calculations. This is motivated by the fact that the results found in Section \ref{sechighv} are identical to the ones 
that can be deduced by taking the $v\to 1$ limit of the results of Section \ref{seclowv}. However, it is interesting to sketch how the computation of 
$\Delta_{11}$ would be carried out in light-cone coordinates for $v\sim 1$. One would start with the NRQED Lagrangian that includes the interaction 
with collinear photons of Eq.~(\ref{lagrascet2}). Since the virtuality of the collinear photons is of order $T^2$, and  $T\gg 1/r$,  they can be integrated out before  evaluating  the potential. At leading order, this gives rise to an energy shift similar to the one reported in Eq.~(\ref{eshift}). 
In the light sector of the NRQED Lagrangian there are also interactions between soft photons and collinear electrons and integrating out  collinear electrons  one obtains the HTL Lagrangian. If the scale $T_-$ is much smaller than $1/r$ one should consider its effects in the ultrasoft photons. However, these photons can only give subleading contributions by means of loop corrections.
From now on we consider that the distinction between moderate and relativistic velocities is not essential and will not be done.

\begin{figure}[htbp!]
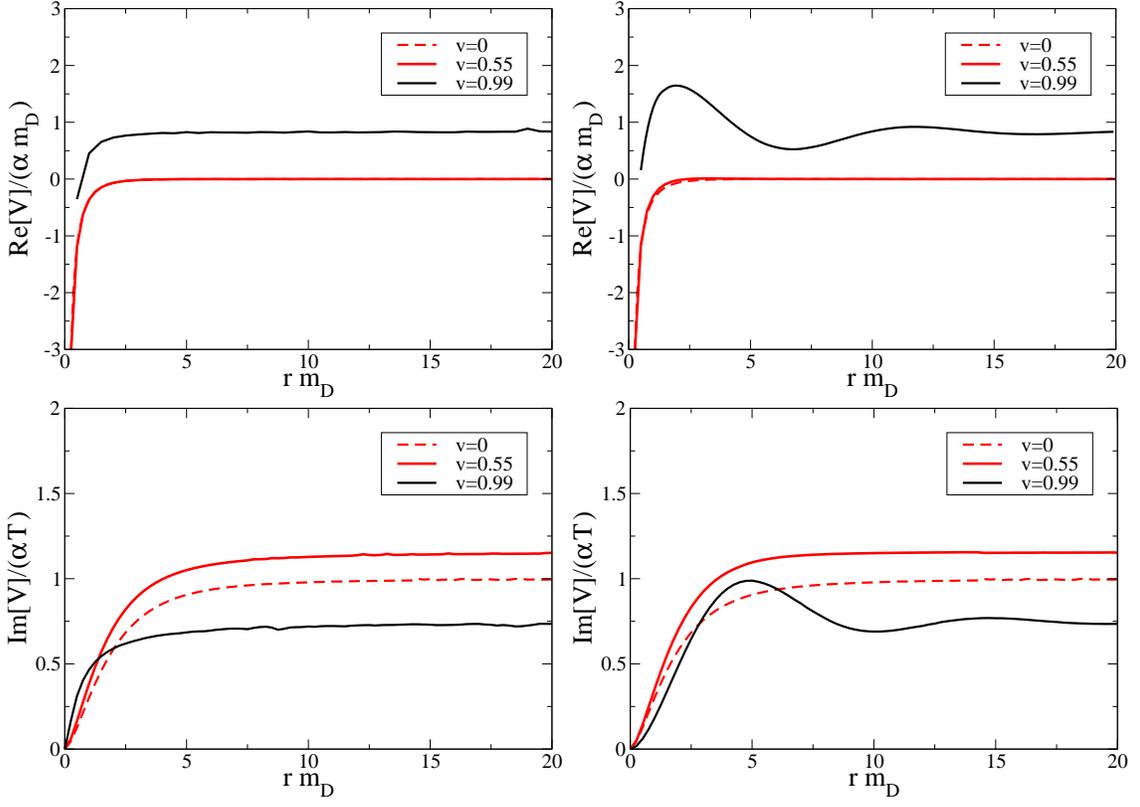

\includegraphics[scale=0.3]{./real-thetapi2.eps}
\includegraphics[scale=0.3]{./real-theta0.eps}
\includegraphics[scale=0.3]{./imaginary-thetapi2.eps}
\includegraphics[scale=0.3]{./imaginary-theta0.eps}
\caption{Real (upper panels) and imaginary (lower panels) parts of the  potential between a quark and an antiquark moving 
with velocities $v=0$, $0.55$, $0.99$ with respect to the thermal bath. 
Right (left) panels correspond to the direction parallel (perpendicular) to the velocity of the thermal medium.}
\label{potential-v055}
\end{figure}

\begin{figure}[htbp!]
\begin{center}
$v=0$\\
\includegraphics[scale=0.6]{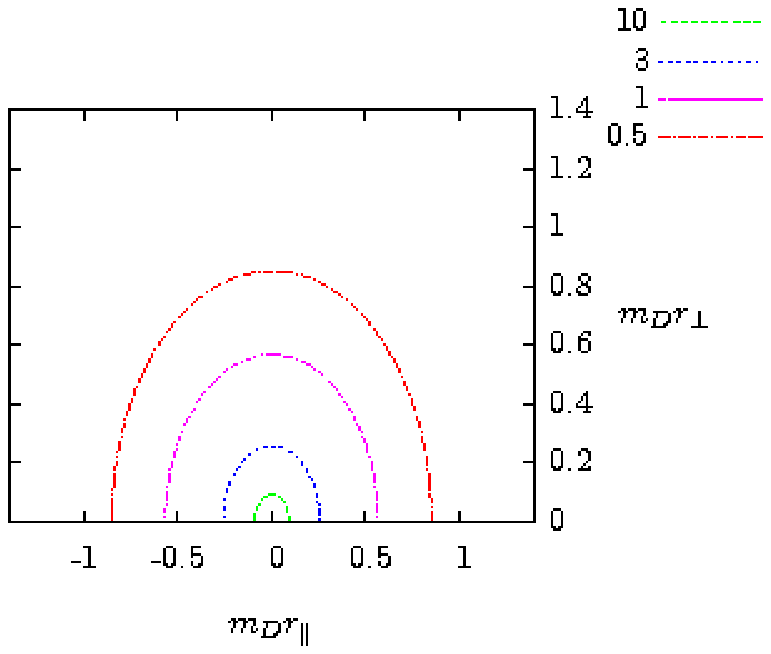}\includegraphics[scale=0.6]{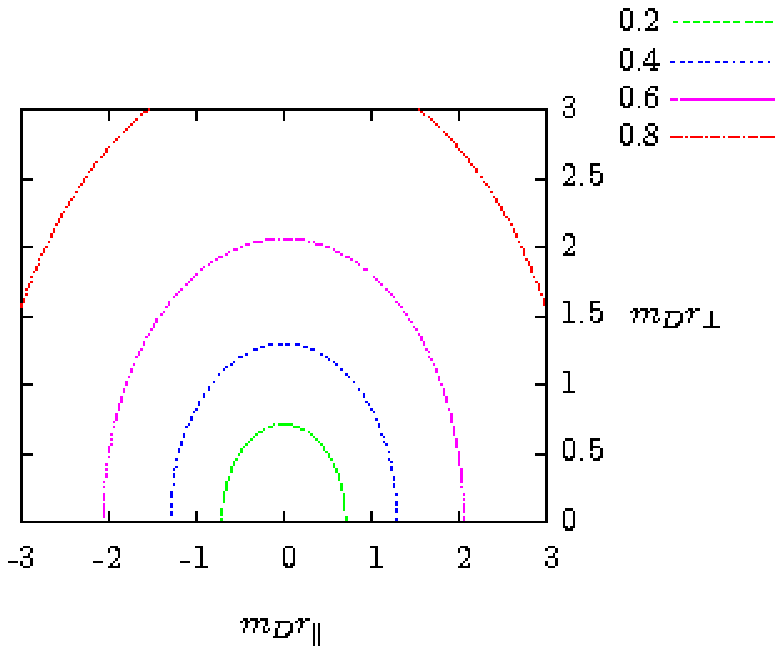}\\
$v=0.5$\\
\includegraphics[scale=0.6]{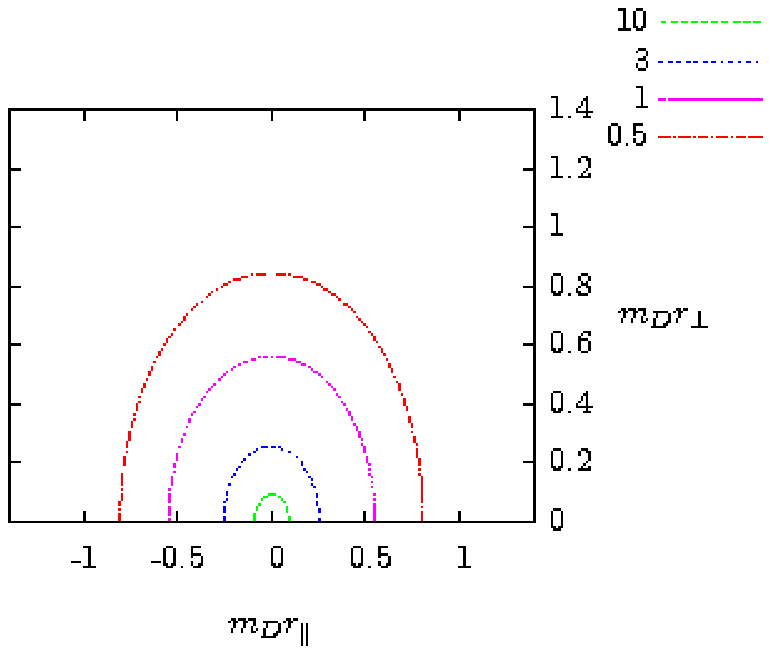}\includegraphics[scale=0.6]{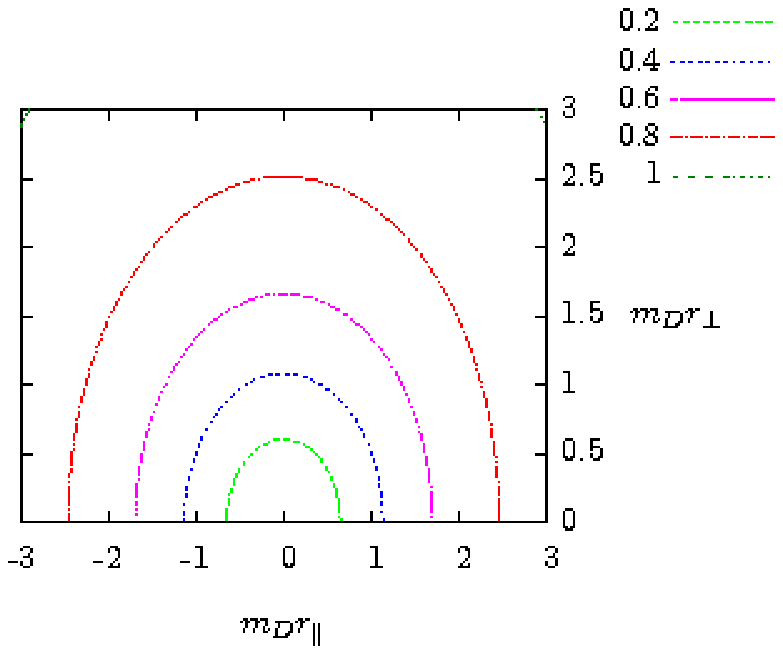}\\
$v=0.9$\\
\includegraphics[scale=0.6]{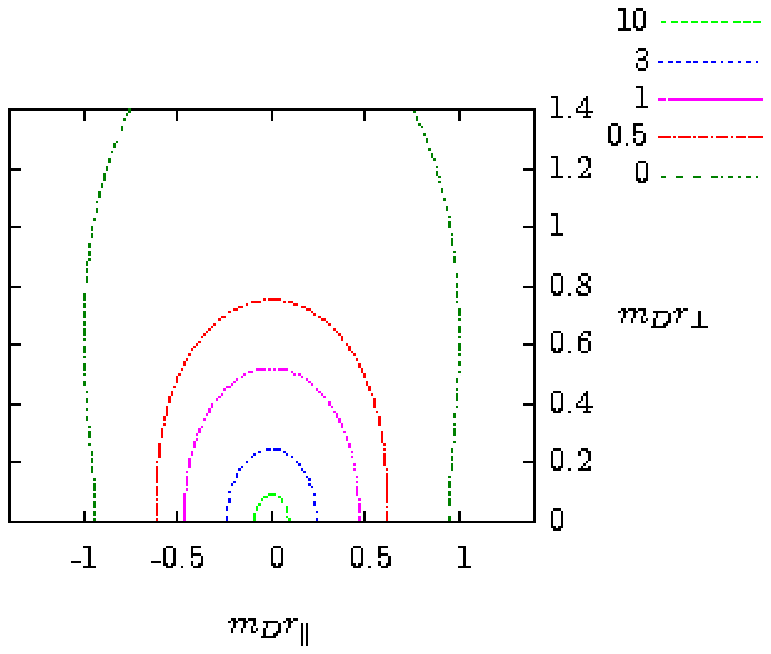}\includegraphics[scale=0.6]{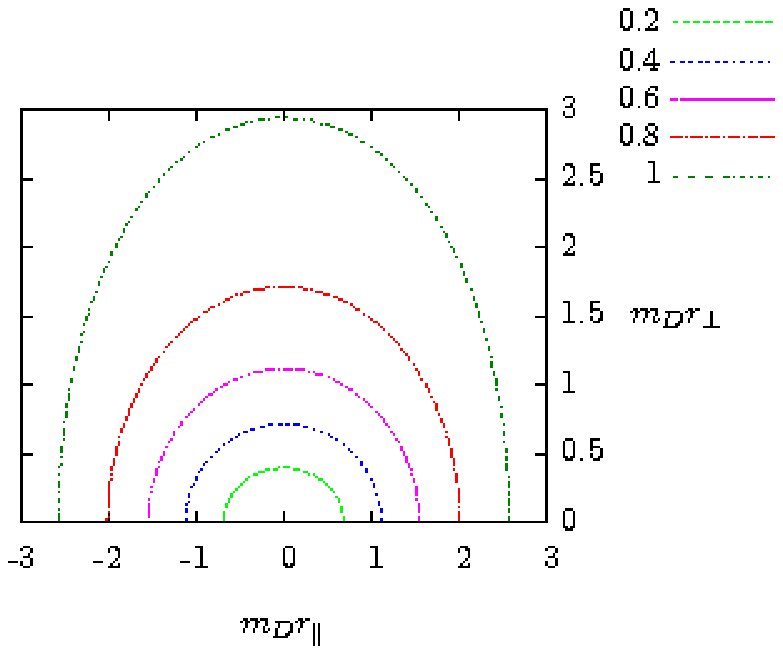}\\
$v=0.99$\\
\includegraphics[scale=0.6]{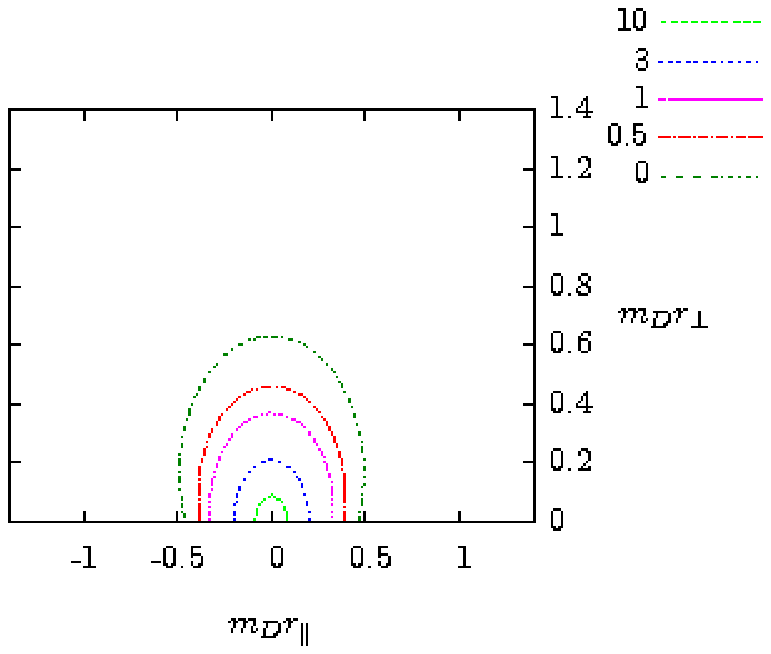}\includegraphics[scale=0.6]{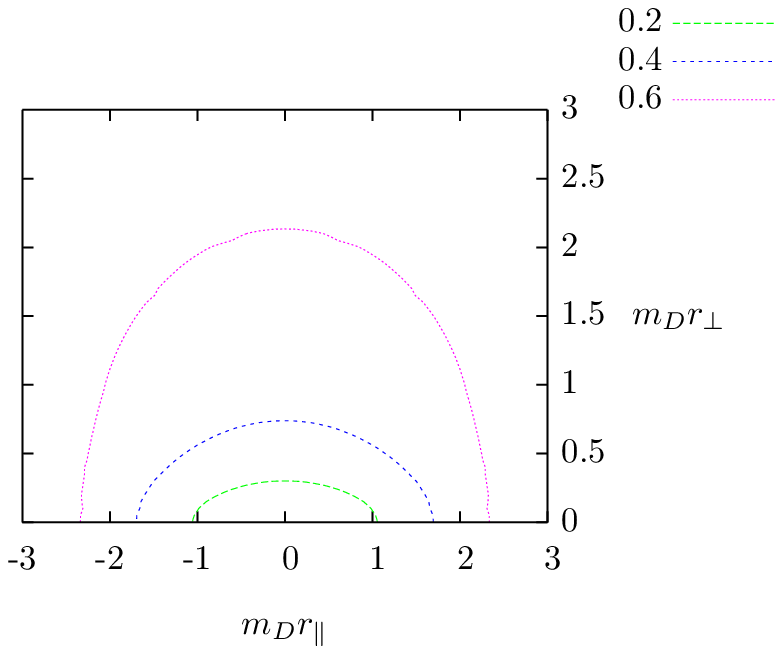}
\caption{Contour lines of the two-body potential between a quark and antiquark moving with respect to the thermal bath with four different values of the velocity: $v=0, 0.5, 0.9, 0.99$. The left panels correspond to the real part of the potential, while the right panels correspond to the imaginary part of the potential. The real part of the potential has been normalized to $\alpha m_D$, while the imaginary part of the potential has been normalized to $\alpha T $.}
\label{potcont}
\end{center}
\end{figure}

From the Fourier transform of the $\Delta_{11}$ propagator  we have determined the real and imaginary parts of the potential in the HTL approximation.
 The potential is anisotropic, and  in Fig.~\ref{potential-v055} 
we display the plots of the real (upper panels) and imaginary (lower panels) part of the potential for $v=0$, $v=0.55$, and $v=0.99$ respectively.
 We consider two directions: the first is along the direction of movement of the thermal bath (right panels) and the second one is  along the 
direction orthogonal to the thermal bath (left panels). 
We plot only positive values of $r$ because the potential is symmetric for $r \to -r$.
We normalize the real part of the potential to  $\alpha  m_D$, which describes the typical strength of the potential in the $v=0$ case. 
The imaginary part of the potential is normalized to $\alpha T$. With these normalizations the displayed shapes hold both for muonic hydrogen and 
for heavy quarkonium.

Regarding the real part,
we observe that for $v\sim 0.55$ it is very similar (in fact, the curves overlap to a large extent) to the real part in the 
$v=0$ case, being  Debye screened at a distance of order $m_D$ and not very asymmetric.  For $v\sim 1$, however, although the real part of the potential 
remains Debye screened at roughly the same distance, it develops a rather large anisotropy.
Indeed, an oscillation is observed in the direction of  motion,  which leads to the formation of a wake in the plasma.

Concerning the imaginary part of the potential for $v\sim 0.55$, it is not very asymmetric and remains very similar 
to the $v=0$ case. It monotonically increases and keeps the same pattern until $v\sim 0.9$. From that velocity on the 
imaginary part decreases and 
the anisotropy grows (see the $v=0.99$ curve). 
In the direction parallel to $\bf v$ one has that there is a mild increase with respect to the $v=0$ case for 
$r \simeq 4 m_D$, and an oscillatory behavior at larger $r$ is also displayed. 
In the direction orthogonal to $\bf v$ there is an enhancement of the potential at short range and a decrease at large distances. 
Note that the imaginary part of the potential vanishes in the origin for any value of $v$ and in any direction.

In Fig.~\ref{potcont} we plot the contour lines for the real and imaginary parts of the two-body potentials between two
 heavy particles with opposite charges, for various values of the velocity. We  focus on the short distance regime 
(the normalization for the real part is slightly different from the one in Fig.~\ref{potential-v055} because we have 
inverted the sign in order to match the normalization of \cite{Chu:1988wh}), because  the distances relevant for
 dissociation are in the range $r m_D  \lesssim 1$ \cite{Escobedo:2008sy}. 
Various plots  of the contour lines of the real part of the two-body
 potential were also shown in \cite{Chu:1988wh} (see also \cite{Chakraborty:2006md}) and it can be seen that we obtained 
exactly the same results. 
The contour lines for the imaginary part of the two-body potential are reported for the first time here, and we observe that an important anisotropy exists even at short distances.

Let us next estimate the dissociation temperature in a way similar to refs. \cite{Escobedo:2008sy,Escobedo:2010tu}. If we 
assume that the 
typical momentum transfer $k$ is larger than the velocity dependent screening mass 
$m_D^2(v,\theta) \sim \vert \Pi_R (k,u)\vert$, we obtain that at $k\sim e^{2\over 3} T\sqrt{1-v^2}$ 
the real and the imaginary parts of the potential have the same size\footnote{This is so for generic $\theta$, 
meaning $\theta \nsim \pi /2$. For $\theta \sim \pi /2$, we obtain $k\sim e^{2\over 3} T/(1-v^2)^{1/3}$.}. For moderate 
velocities no qualitative change is expected with respect to the $v=0$ case. However, for $v$ close 
to $1$, $k$ becomes small and it is not guaranteed that the screening mass can be neglected in front of it. Indeed,
we find that the screening mass remains finite for $v$ close to $1$ \footnote{This is so for generic $\theta$, 
meaning $\theta \nsim \pi /2$. For $\theta \sim \pi /2$, we obtain $m_D^2(v,\theta) \sim m_D^2/(1-v^2)$.}. This means that 
the typical $k$ for which the real and imaginary parts of the potential have the same size is smaller than the 
screening mass. Since a screened 
potential only supports bound states of a typical $k$ larger than $m_D(v,\theta )$, we conclude that at relativistic 
velocities, unlike the case of moderate velocities, the dissociation occurs due to screening 
(i.e. at the scale $T_d\sim m_e e$), as originally proposed
by Matsui and Satz \cite{Matsui:1986dk}, rather than due to Landau damping \cite{Laine:2006ns,Escobedo:2008sy,Brambilla:2008cx}. 
This can also be qualitatively understood from our plots.
For $v$ large and increasing, we see from Fig. \ref{potential-v055} that the real part of the potential increases whereas the
imaginary part decreases. Therefore, from some $v$ on, the real part of the potential dominates over the imaginary part 
and one can find the approximate wave functions of the system by solving a standard Schr\"odinger equation with a 
real potential. The decay width may be then calculated in perturbation theory by sandwiching the 
imaginary part of the potential between those wave functions. The wave functions of bound states
go to a vanishing value at the distance where the real part of the potential becomes flat.  From Fig. \ref{potcont} it is 
also clear that the real part of the potential at 
short distances becomes steeper at increasing $v$. On the one hand this implies that no bound state exists from a certain 
velocity on. On the other hand it implies that when bound states still exist their wave functions are increasingly
localized close to the origin. Since the imaginary part of the potential goes to zero at the origin, 
it follows that the decay width of such states is also going to zero at increasing $v$. 

$v_c$ at which screening overtakes Landau damping as the dominant mechanism for dissociation, by equating $e T$ to
$e^{2/3} T (1-v^2)^{1/2}$ above. We obtain $v_c \sim \sqrt{1-ae^{2/3}}$, where $a$ is a numerical factor of order one
\footnote{This is so for generic $\theta$, 
meaning $\theta \nsim \pi /2$. For $\theta \sim \pi /2$, we obtain $v_c \sim \sqrt{1-ae^{2}}$.}.
A quantitative study of all these issues may be carried out by numerically solving the Schr\"odinger equation with the full
(complex) potential, for instance along the lines \cite{Burnier:2007qm,Miao:2010tk,Margotta:2011ta}. This is however beyond 
the scope of this paper.

Regarding the real part of the potential, we find it interesting to compare the results reported in 
Fig.~\ref{potential-v055} 
with the recent results obtained for super Yang-Mills theory using AdS/CFT~\cite{Liu:2006nn}. For this theory it was 
stated that the potential could 
be approximated with a Yukawa potential and the dependence on the velocity encoded in a screening length that depends 
on $v$ and $\theta$ as follows:
\be
m_D(v,\theta)=m_D(0,0)\frac{h(v,\theta)}{(1-v^2)^{1/4}} \,,
\label{ads}
\ee
where $h(v,\theta)$ is a function that is almost constant for any $v$ and $\theta$. 
The expression above does not give a good approximation of the potential in the HTL approximation. In particular the Debye 
screening for $v=0.99$ is 
strongly dependent on the angle $\theta$. If we try to fit the exponential behavior of the longitudinal and transverse 
directions, we find a screening
length which is about a factor $2$ larger in the longitudinal direction with respect to the transverse direction.  
This can also be inferred from the fact that the scale of $m_D^2(v,\theta)$ must be given by $\vert \Pi_R (k,u)\vert$.
 We obtain in the case $v\to 1$
\be
m_D(v,\theta)\sim \left\{
\begin{array}{ll}
m_D\vert \tan \theta\vert & {\rm if} \theta\nsim {\pi\over 2}\\ & \\
{m_D\over \sqrt{1-v^2}}  & {\rm if} \theta \sim {\pi\over 2} 
\end{array}\right.
\label{mDvto1}
\ee
Be aware that the $\theta$ above is the angle between the velocity and the momentum transfer, whereas the $\theta$ in 
(\ref{ads}) is the angle between the velocity and the relative position. In any case, the expression 
(\ref{mDvto1}) shows that at ultrarelativistic velocities a strong anisotropy for real space potential is expected, as 
confirmed by our figures and discussed above. 
 
Regarding  the oscillatory part of the potential one might wonder  whether it is due to the weak coupling approximation. However, as shown in~\cite{Jiang:2010zza} for the potential produced by a 
single charge, in the HTL resummation approach the oscillations  of the potential are larger than in the HTL approximation.  Moreover, one would naively expect that with increasing coupling the wakes should be larger than in the weak coupling approximation.

\section{Conclusions}
\label{secconclusions}
The EFT theory for the description of bound states in a thermal medium has several interesting aspects. When the bound 
state moves with a moderate speed
 with respect to the medium the resulting EFT is quite similar to the one developed for the bound state at rest. 
We have taken into account the suitable modifications in Section~\ref{seclowv}, for the hydrogen atom
in the cases $T\ll 1/r$ and $T\sim 1/r$. However, when the speed 
is close to $1$, one has to consider two well separated scales, $T_+$ and $T_-$, defined in Eq.~(\ref{temperatures}), and 
in the corresponding EFT one
 has collinear as well as soft degrees of freedom. The effective temperatures $T_+$ and $T_-$  can be in two different 
energy ranges and  in Section~\ref{sechighv}
 we have considered two specific cases: the first one corresponds to  $T_+\sim 1/r\gg T_-\gg E$ and the second one 
corresponds to 
$T_+\sim m_e\gg 1/r\gg T_-\gg E$. Note that in this case large logarithms of $T_-/T_+$ appear in the calculation. 
The factorized results displayed 
in Appendix \ref{apen2} may be useful for a resummation of these large logarithms. It is reassuring that our results 
for moderate velocities 
are able to reproduce the 
ones obtained for the $v\sim 1$ case. For all the cases above we observe that the thermal decay width monotonically 
decreases with the velocity. This means 
that the faster the bound state moves across the thermal bath the more stable it becomes.

Finally, in Section \ref{secphoton} we have considered the case $T\gg 1/r$ allowing for light fermion pairs in the thermal
 bath.
In atomic physics this state could be the muonic hydrogen in a thermal bath of electrons and positrons, while in heavy-ion
 collisions it may represent heavy quarkonia in the quark-gluon plasma at very high temperatures. We have  determined  
how the imaginary and real component of the two-body potential are modified for nonvanishing velocities of the bound state
 with respect to the medium. 
Regarding the real part of the potential we have reproduced known results, and extended them to higher speeds. 
The imaginary part has been calculated for the first time. Its behavior is similar to 
the one determined for the thermal bath at rest for moderate velocities, but it tends to zero at 
 velocities close to $1$. This implies that Landau damping \cite{Laine:2006ns,Escobedo:2008sy,Brambilla:2008cx} is not the relevant mechanism for dissociation of bound 
states from a certain critical velocity $v_c$ on, which has been estimated in the previous section. Screening, 
as originally proposed by Matsui and Satz \cite{Matsui:1986dk}, becomes then the relevant mechanism. Our results 
for the thermal decay width disagree with the qualitative estimate of ref. \cite{Dominguez:2008be}, and with 
the more quantitative one of ref. \cite{Song:2007gm}. We believe that the main reason for the discrepancy is due to the fact that the velocity dependence of the interaction is not properly taken into
account in those works. Note that our results for the imaginary part depend crucially on the use of the correct non-equilibrium expression in Eq. (\ref{noeq}), which leads to  Eq. (\ref{Pis00}).

In the present paper we have paved the way for a more detailed study of the propagation of bound states in a thermal
 medium. We have assumed that the
 medium is a weakly coupled plasma, moving homogeneously and at a constant temperature, therefore our study needs a number of refinements to be
 realistically applied to HQ states in heavy-ion collisions. In that case, one should consider the expansion and cooling 
of the thermal medium, as well as possible anisotropies 
\cite{Burnier:2009yu,Dumitru:2009fy,Philipsen:2009wg,Chandra:2010xg}. In any case, we expect that the qualitative features
we observe, namely that the decay width decreases with increasing velocity, and hence that Landau damping ceases to be 
the relevant mechanism for dissociation at a certain critical velocity, will remain true.    

\acknowledgments

We thank Cristina Manuel for discussions in the early stages of this work, and Jos\'e Mar\'\i a Fern\'andez Varea 
for explanations on the wake and for bringing to our attention 
ref. \cite{cond-mat}. 
JS also thanks David d'Enterria for useful communications.
MM and JS have been supported by the CPAN  CSD2007-00042 Consolider-Ingenio 2010 program and the 2009SGR502 CUR grant 
(Catalonia).  MM has also been supported by the grant FPA2007-66665-C02-01(Spain). MAE and JS have been suported by the 
RTN Flavianet MRTN-CT-2006-035482 (EU), and the FPA2007-60275 and FPA2010-16963 projects (Spain). 
MAE has also been supported by a MEC FPU fellowship (Spain).

\appendix
\section{Computation of $A$ and $B$ from section \ref{seclowv}}
\label{apen1}
As a starting point one can use Eqs.~(16) and (17) from~\cite{Escobedo:2008sy}, which for a bound state in a static thermal bath give 
\begin{equation}
I_{ii}(q)=\frac{q}{2\pi^2}\left(\log\left(\frac{2\pi T}{|q|}\right)+\Re\Psi\left(\frac{i|q|}{2\pi T}\right)\right)\,.
\end{equation}
This equation is obtained by a trivial angular integration because the system is symmetric under space rotations. When the bound state moves with respect to the thermal bath,  the distribution function depends on the effective temperature defined in Eq.~(\ref{effective-temperature}) and has a nontrivial dependence of the angle between $\bf k$ and $\bf v$. Now if we take into account the structure that was shown in Eq.~(\ref{relA}) and define $x=\cos \theta$, we have that
\begin{equation}
\Im A(q)=\frac{q}{8\pi^2}\int_{-1}^1\,dx\left(\log\left(\frac{2\pi T\sqrt{1-v^2}}{|q|(1-vx)}\right)+\Re\Psi\left(\frac{i|q|(1-vx)}{2\pi T\sqrt{1-v^2}}\right)\right)\,,
\end{equation}
and  for the real part
\begin{equation}
\Re A(q)=\frac{1}{8\pi}\int_{-1}^1\,dx\frac{|q|}{e^{\frac{|q|(1-vx)}{T\sqrt{1-v^2}}}-1}\,.
\end{equation}
Using Eq.~(\ref{relA}) and the above equations we obtain that the imaginary and real part of the coefficient $B$ are respectively given by
\begin{equation}
\Im B(q)=\frac{q}{8\pi^2}\int_{-1}^1\,dx x^2\left(\log\left(\frac{2\pi T\sqrt{1-v^2}}{|q|(1-vx)}\right)+\Re\Psi\left(\frac{i|q|(1-vx)}{2\pi T\sqrt{1-v^2}}\right)\right)\,,
\end{equation}
and
\begin{equation}
\Re B(q)=\frac{1}{8\pi}\int_{-1}^1\,dxx^2\frac{|q|}{e^{\frac{|q|(1-vx)}{2\pi T\sqrt{1-v^2}}}-1}\,.
\end{equation}
As a cross-check, in the $v=0$ limit we find that the relation $B=\frac{A}{D-1}$ is fulfilled, and combining this with the identity
\begin{equation}
\delta_{ij}=\frac{3}{2}P^s_{ij}+\frac{1}{2}P^p_{ij}\,,
\end{equation}
one recovers the results  reported in~\cite{Escobedo:2008sy}.

\section{Computation of the contribution from ultrasoft photons in section \ref{sechighv}}
\label{apen2}
In this appendix we compute the matrix elements of $K_{ij}$ defined in Eq.~(\ref{uscon}) in the various integration regions identified in Section~\ref{sechighv}.

\subsection{The $k_+,k_-\sim T_-$ region}
The quantities $a$ and $b$ defined in Eq.~(\ref{defK})   can be computed from $K_{ij}$ as follows
\begin{eqnarray}
a &=&\frac{2}{2D-4}K_{ii}+\frac{D-4}{2D-4}\frac{v^iv^j}{v^2}K_{ij}\,, \\
b &=&\frac{2}{2D-4}K_{ii}-\frac{D}{2D-4}\frac{v^iv^i}{v^2}K_{ij}\,,
\end{eqnarray}
and for $k_+,k_-\sim T_-$ we find that
\begin{equation}
K_{ij}(q)=-2i\int\frac{\,d^Dk}{(2\pi)^{D-1}}\frac{\delta(k_+k_--k_\perp^2)}{e^{\frac{|k_-|}{2T_-}}-1}\left(\delta_{ij}-\frac{k_ik_j}{k^2}\right)\left[\frac{1}{k_++k_--i\epsilon}+\frac{2q}{(k_++k_--i\epsilon)^2}+...\right]\,.
\end{equation}
The first term in the square brackets vanishes by symmetry considerations and we find that
\begin{eqnarray}
\Re K_{ii}(q)&=&0 \,,\\
 \Im K_{ii}(q)&=&\frac{q}{2\pi^2}\left(\frac{1}{D-4}+\frac{1}{2}-\frac{\gamma}{2}+\frac{1}{2}\log\pi+\log\left(\frac{2T_-}{\mu}\right)\right) \,,
\end{eqnarray}
and 
\begin{eqnarray}
\Re\frac{v^iv^j}{v^2}K_{ij}(q)&=& 0\,, \\
\Im\frac{v^iv^j}{v^2}K_{ij}(q)&=& \frac{q}{6\pi^2}\left(\frac{1}{D-4}-\frac{\gamma}{2}+\frac{1}{2}\log\pi+\log\left(\frac{2T_-}{\mu}\right)\right)\,.
\end{eqnarray}

\subsection{The case with $k_+,k_-\sim q$ }
In this region we have that
\begin{equation}
K_{ij}(q)=i\int\frac{\,d^Dk}{(2\pi)^{D-1}}\delta(k_+k_--k_\perp^2)\left(\frac{2T_-}{|k_-|}-\frac{1}{2}+...\right)\frac{1}{q-\frac{1}{2}(k_++k_-)+i\epsilon}\,,	
\end{equation}
and it is useful to calculate separately the imaginary part of the  integrals with the first  and the second terms in the brackets. The reason is that the computation of the first  term in dimensional regularization is technically difficult, while the second one is quite straightforward. 

We first compute the imaginary part of the term lineal in $T_-$ with a cut-off to separate  the region $k_+\sim q$ from the region with $k_-\sim q(T_-/T_+)$. Thus we consider a cut-off $\Lambda$ such that $q\gg \Lambda\gg q(T_-/T_+)$ and we obtain
\be
\Im K_{ii}(q)=2T_-\int_0^\infty\,dk_+\left[\frac{\log\left(\frac{\Lambda}{k_++2q+i\epsilon}\right)}{k_++2q+i\epsilon}+\frac{\log\left(\frac{\Lambda}{k_++2q-i\epsilon}\right)}{k_++2q-i\epsilon}-\frac{\log\left(\frac{\Lambda}{k_+-2q+i\epsilon}\right)}{k_+-2q+i\epsilon}-\frac{\log\left(\frac{\Lambda}{k_+-2q-i\epsilon}\right)}{k_+-2q-i\epsilon}\right]\,,
\ee
and 
\be
\Im\frac{v^iv^j}{v^2}K_{ij}(q)=0\,.
\ee
Then, the remaining  terms are computed in dimensional regularization. Summing all the terms we obtain that
\begin{eqnarray}
\Re K_{ii}(q)&=&\frac{T_-}{\pi}\left(\frac{1}{D-4}+\frac{1}{2}+\frac{\gamma}{2}-\frac{1}{2}\log\pi+\log\left(\frac{|q|}{\mu}\right)\right)-\frac{|q|}{4\pi}\,, \\
\Im K_{ii}(q) &=&-\frac{q}{2\pi^2}\left(\frac{1}{D-4}-\frac{1}{2}\log\pi+\log\left(\frac{|q|}{\mu}\right)-\frac{1}{2}+\frac{\gamma}{2}\right) \nonumber \\
&+&2T_-\int_0^\infty\,dk_+\left[\frac{\log\left(\frac{\Lambda}{k_++2q+i\epsilon}\right)}{k_++2q+i\epsilon}+\frac{\log\left(\frac{\Lambda}{k_++2q-i\epsilon}\right)}{k_++2q-i\epsilon}-\frac{\log\left(\frac{\Lambda}{k_+-2q+i\epsilon}\right)}{k_+-2q+i\epsilon}-\frac{\log\left(\frac{\Lambda}{k_+-2q-i\epsilon}\right)}{k_+-2q-i\epsilon}\right]\,, 
\end{eqnarray}
and
\begin{eqnarray}
\frac{v^iv^j}{v^2} \Re K_{ij}(q)&=&\frac{T_-}{2\pi}-\frac{|q|}{6\pi} \,,\\
\frac{v^iv^j}{v^2}\Im K_{ij}(q)&=&-\frac{q}{6\pi^2}\left(\frac{1}{D-4}-\frac{5}{6}+\frac{\gamma}{2}-\frac{1}{2}\log\pi+\log\left(\frac{|q|}{\mu}\right)\right)\,.
\end{eqnarray}

\subsection{The case with $k_+\sim q$, $k_-\sim q(T_-/T_+)$}
In this region we have that
\begin{equation}
K_{ij}(q)=2i\int\frac{\,d^Dk}{(2\pi)^{D-1}}\frac{\delta(k_+k_--k_\perp^2)}{\left|\frac{k_+}{T_+}+\frac{k_-}{T_-}\right|}\left(\delta_{ij}-\frac{k_ik_j}{k^2}\right)\frac{1}{q-\frac{k_+}{2}+i\epsilon}\,.
\end{equation}
For simplicity, we compute the imaginary part using a cut-off (as we did in the previous subsection) and the real part using dimensional regularization. We obtain that the  real and imaginary parts of the trace of $K_{ij}$ are, respectively, given by
\begin{eqnarray}
\Re K_{ii}(q)&=&-\frac{T_-}{\pi}\left(\frac{1}{D-4}+\frac{1}{2}+\frac{\gamma}{2}-\frac{1}{2}\log\pi+\log\left(\frac{|q|}{\mu}\right)+\frac{1}{2}\log\left(\frac{T_-}{T_+}\right)\right) \,,\\
\Im K_{ii}(q)&=&\frac{T_-}{2\pi^2}\int_0^\infty\,dk_+\left[\frac{P}{k_++2q}-\frac{P}{k_+-2q}\right]\log\left(\frac{\Lambda T_+}{k_+T_-}\right)\,,
\end{eqnarray}
where $P$ stands for principal value. Moreover, we find  that
\begin{eqnarray}
\frac{v^iv^j}{v^2}\Re K_{ij}(q)&=&0 \,,\\
\frac{v^iv^j}{v^2}\Im K_{ij}(q)&=&0\,.
\end{eqnarray}

\section{Interaction with collinear photons}
In this appendix we give the detailed form of the part of the EFT's Lagrangians that describes the interaction of electrons with collinear photons. General remarks about the matching are given in the corresponding subsections. 
\subsection{pNRQED Lagrangian for $T_+\sim 1/r$ and $1/r\gg T_-\gg E$}
\label{ape:pnrqed}
In this case the pNRQED Lagrangian has the form
{\small \begin{eqnarray}
\label{lagrascet1}
\delta\mathcal{L}_{pNRQED}&=&c_1\frac{\psi^\dagger\psi}{m_e}\frac{\bar{n}^\mu F_{\mu i}}{(\bar{n}\partial)}\frac{\bar{n}^\nu F_{\nu i}}{(\bar{n}\partial)}+c_2\frac{\psi^\dagger\psi}{m_e}\frac{\bar{n}^\mu n^\nu F_{\mu\nu}}{(\bar{n}\partial)}\frac{\bar{n}^\alpha n^\beta F_{\alpha\beta}}{(\bar{n}\partial)}+c_3\frac{\psi^\dagger\psi}{m_e}\left[\frac{n^\mu F_{i\mu}}{(\bar{n}\partial)}\frac{\bar{n}^\nu F_{\nu i}}{(\bar{n}\partial)}+\frac{\bar{n}^\mu F_{i\mu}}{(\bar{n}\partial)}\frac{n^\nu F_{\nu i}}{(\bar{n}\partial)}\right]\nonumber \\
&+&c_4\frac{\psi^\dagger\psi}{m_e}\frac{n^\mu F_{i\mu}}{(\bar{n}\partial)}\frac{n^\nu F_{i\nu}}{(\bar{n}\partial)}+c_5\frac{\psi^\dagger\psi}{m_e}\left[\frac{\bar{n}^\mu n\partial F_{\mu i}}{(\bar{n}\partial)^2}\frac{\bar{n}^\nu F_{\nu i}}{(\bar{n}\partial)}+\frac{\bar{n}^\mu F_{\mu i}}{(\bar{n}\partial)}\frac{\bar{n}^\nu n\partial F_{\nu i}}{(\bar{n}\partial)^2}\right]\nonumber \\
&+&c_{6}\frac{\psi^\dagger\psi}{m_e}\left[\frac{\bar{n}^\mu n^\nu n\partial F_{\mu\nu}}{(\bar{n}\partial)^2}\frac{\bar{n}^\alpha n^\beta F_{\alpha\beta}}{(\bar{n}\partial)}+\frac{\bar{n}^\mu n^\nu F_{\mu\nu}}{(\bar{n}\partial)}\frac{\bar{n}^\alpha n^\beta n\partial F_{\alpha\beta}}{(\bar{n}\partial)^2}\right] \nonumber \\
&+&\frac{ic_7}{m_e^2}\left[\psi^\dagger\frac{\bar{n}^\mu n^\nu F_{\mu\nu}}{(\bar{n}\partial)}\frac{\bar{n}^\alpha n^\beta F_{\alpha\beta}}{(\bar{n}\partial)}D_3\psi-D_3\psi^\dagger\frac{\bar{n}^\mu n^\nu F_{\mu\nu}}{(\bar{n}\partial)}\frac{\bar{n}^\alpha n^\beta F_{\alpha\beta}}{(\bar{n}\partial)}\psi\right]\nonumber \\
&+&\frac{ic_8}{m_e^2}\left(\psi^\dagger\left[\frac{\bar{n}^\mu n^\nu F_{\mu\nu}}{(\bar{n}\partial)}\frac{\bar{n}^\alpha F_{\alpha i}}{(\bar{n}\partial)}+\frac{\bar{n}^\mu F_{\mu i}}{(\bar{n}\partial)}\frac{\bar{n}^\alpha n^\beta F_{\alpha\beta}}{(\bar{n}\partial)}\right]D_i\psi-D_i\psi^\dagger\left[\frac{\bar{n}^\mu n^\nu F_{\mu\nu}}{(\bar{n}\partial)}\frac{\bar{n}^\alpha F_{\alpha i}}{(\bar{n}\partial)}+\frac{\bar{n}^\mu F_{\mu i}}{(\bar{n}\partial)}\frac{\bar{n}^\alpha n^\beta F_{\alpha\beta}}{(\bar{n}\partial)}\right]\psi\right)\nonumber \\
&+&\frac{ic_9}{m_e^2}\left(D_i\psi^\dagger\left[\frac{\bar{n}^\mu F_{\mu j}}{(\bar{n}\partial)}\frac{\bar{n}^\nu\partial_jF_{\nu i}}{(\bar{n}\partial)^2}+\frac{\bar{n}^\mu\partial_jF_{\mu j}}{(\bar{n}\partial)^2}\frac{\bar{n}^\nu F_{\nu j}}{(\bar{n}\partial)}\right]\psi-\psi^\dagger\left[\frac{\bar{n}^\mu F_{\mu j}}{(\bar{n}\partial)}\frac{\bar{n}^\nu\partial_jF_{\nu i}}{(\bar{n}\partial)^2}+\frac{\bar{n}^\mu\partial_jF_{\mu j}}{(\bar{n}\partial)^2}\frac{\bar{n}^\nu F_{\nu j}}{(\bar{n}\partial)}\right]D_i\psi\right)\nonumber \\
&+&\frac{ic_{10}}{m_e^2}\left(D_i\psi^\dagger\left[\frac{n^\mu F_{i\mu}}{(\bar{n}\partial)}\frac{\bar{n}^\nu n^\alpha F_{\nu\alpha}}{(\bar{n}\partial)}+\frac{\bar{n}^\mu n^\nu F_{\mu\nu}}{(\bar{n}\partial)}\frac{n^\alpha F_{i\alpha}}{(\bar{n}\partial)}\right]\psi-\psi^\dagger\left[\frac{n^\mu F_{i\mu}}{(\bar{n}\partial)}\frac{\bar{n}^\nu n^\alpha F_{\nu\alpha}}{(\bar{n}\partial)}+\frac{\bar{n}^\mu n^\nu F_{\mu\nu}}{(\bar{n}\partial)}\frac{n^\alpha F_{i\alpha}}{(\bar{n}\partial)}\right]D_i\psi\right) \,.
\end{eqnarray}}
The $\psi$ field is the field of the electron in pNRQED (we are using the form of pNRQED shown in Eq.~(3) of \cite{Pineda:1997ie}), 
$D_i$ is the covariant derivative (containing ultrasoft photons only) that acts on the field of the electron and Latin indices stand always for transverse components. 

The power counting for this Lagrangian works as follows. The leading order terms are the ones proportional to $c_1$ and $c_2$; note that the two terms in $\bar{n}^\mu F_{\mu i}$ are not of the same order of magnitude as $\bar{n}\partial A_i\gg\partial_i\bar{n}A$. Taking only the leading term of $F_{\mu\nu}$ in this two terms they have the form
\begin{equation}
\label{eq:c1c2}
c_1\frac{\psi^\dagger\psi A_\perp^2}{m_e}+c_2\frac{\psi^\dagger\psi(nA)^2}{m_e},
\end{equation}
we consider that all the components of $A_\mu$ are of the same approximate size (this may not be true in some specific 
gauges). Starting from (\ref{eq:c1c2}) each term that has an additional covariant derivative acting on an electron field
 is suppressed by an order $\alpha$, each transverse derivative acting on $A_\mu$ suppresses this term by an order 
$\sqrt{\frac{1-v}{1+v}}$ and each $(n\partial)$ acting on a $A_\mu$ field suppresses the term by an order 
$\frac{1-v}{1+v}$. If we consider that $\frac{1-v}{1+v}\gg\alpha\gg\left(\frac{1-v}{1+v}\right)^{3/2}$ the terms shown 
in Eq.~(\ref{lagrascet1}) provide the complete list of operators
 at the  order $m_e\alpha^5$. In practice only the two first terms contribute to our calculation. The Wilson coefficients 
are fixed by the matching calculation schematically shown in~(\ref{match1}). We obtain
\begin{eqnarray}
c_1&=&c_2=c_4=c_9=\frac{e^2}{2}\,, \\
c_3&=&c_7=c_6=c_{10}=-\frac{e^2}{2}\,, \\
c_{8}&=&c_5=\frac{e^2}{4}\,.
\end{eqnarray}

\subsection{NRQED Lagrangian for $T_+\sim m_e$ and $1/r\gg T_-\gg E$}
\label{ape:nrqed}
The full expression for the part of the NRQED Lagrangian that deals with the interaction of electrons with collinear photons is
{\small 
\begin{eqnarray}
\label{lagrascet2}
\delta\mathcal{L}_{NRQED}&=&c_1\frac{\psi^\dagger\psi}{m_e}\frac{\bar{n}^\mu F_{\mu i}}{(\bar{n}\partial)}\frac{\bar{n}^\nu F_{\nu i}}{(\bar{n}\partial)}+c_2\frac{\psi^\dagger\psi}{m_e}\frac{\bar{n}^\mu n^\nu F_{\mu\nu}}{(\bar{n}\partial)}\frac{\bar{n}^\alpha n^\beta F_{\alpha\beta}}{(\bar{n}\partial)}+c_3\frac{\psi^\dagger\psi}{m_e}\left[\frac{n^\mu F_{i\mu}}{(\bar{n}\partial)}\frac{\bar{n}^\nu F_{\nu i}}{(\bar{n}\partial)}+\frac{\bar{n}^\mu F_{i\mu}}{(\bar{n}\partial)}\frac{n^\nu F_{\nu i}}{(\bar{n}\partial)}\right]\nonumber \\
&+&\frac{ic_4}{m_e^2}\left[\psi^\dagger\frac{\bar{n}^\mu n^\nu F_{\mu\nu}}{(\bar{n}\partial)}\frac{\bar{n}^\alpha n^\beta F_{\alpha\beta}}{(\bar{n}\partial)}D_3\psi-D_3\psi^\dagger\frac{\bar{n}^\mu n^\nu F_{\mu\nu}}{(\bar{n}\partial)}\frac{\bar{n}^\alpha n^\beta F_{\alpha\beta}}{(\bar{n}\partial)}\psi\right]\nonumber \\
&+&\frac{ic_5}{m_e^2}\left(\psi^\dagger\left[\frac{\bar{n}^\mu n^\nu F_{\mu\nu}}{(\bar{n}\partial)}\frac{\bar{n}^\alpha F_{\alpha i}}{(\bar{n}\partial)}+\frac{\bar{n}^\mu F_{\mu i}}{(\bar{n}\partial)}\frac{\bar{n}^\alpha n^\beta F_{\alpha\beta}}{(\bar{n}\partial)}\right]D_i\psi-D_i\psi^\dagger\left[\frac{\bar{n}^\mu n^\nu F_{\mu\nu}}{(\bar{n}\partial)}\frac{\bar{n}^\alpha F_{\alpha i}}{(\bar{n}\partial)}+\frac{\bar{n}^\mu F_{\mu i}}{(\bar{n}\partial)}\frac{\bar{n}^\alpha n^\beta F_{\alpha\beta}}{(\bar{n}\partial)}\right]\psi\right)\nonumber \\
&+&c_6\frac{\psi^\dagger\psi}{m_e}\frac{n^\mu F_{i\mu}}{(\bar{n}\partial)}\frac{n^\nu F_{i\nu}}{(\bar{n}\partial)}+c_7\frac{\psi^\dagger\psi}{m_e}\left[\frac{\bar{n}^\mu n\partial F_{\mu i}}{(\bar{n}\partial)^2}\frac{\bar{n}^\nu F_{\nu i}}{(\bar{n}\partial)}+\frac{\bar{n}^\mu F_{\mu i}}{(\bar{n}\partial)}\frac{\bar{n}^\nu n\partial F_{\nu i}}{(\bar{n}\partial)^2}\right]\nonumber \\
&+& c_{8}\frac{\psi^\dagger\psi}{m_e}\left[\frac{\bar{n}^\mu n^\nu n\partial F_{\mu\nu}}{(\bar{n}\partial)^2}\frac{\bar{n}^\alpha n^\beta F_{\alpha\beta}}{(\bar{n}\partial)}+\frac{\bar{n}^\mu n^\nu F_{\mu\nu}}{(\bar{n}\partial)}\frac{\bar{n}^\alpha n^\beta n\partial F_{\alpha\beta}}{(\bar{n}\partial)^2}\right]+\frac{ic_9}{m_e^2}\left(D_i\psi^\dagger\left[\frac{\bar{n}^\mu F_{\mu j}}{(\bar{n}\partial)}\frac{\bar{n}^\nu\partial_jF_{\nu i}}{(\bar{n}\partial)^2}+\frac{\bar{n}^\mu\partial_jF_{\mu j}}{(\bar{n}\partial)^2}\frac{\bar{n}^\nu F_{\nu j}}{(\bar{n}\partial)}\right]\psi\right. \nonumber \\
&-&\left.\psi^\dagger\left[\frac{\bar{n}^\mu F_{\mu j}}{(\bar{n}\partial)}\frac{\bar{n}^\nu\partial_jF_{\nu i}}{(\bar{n}\partial)^2}+\frac{\bar{n}^\mu\partial_jF_{\mu j}}{(\bar{n}\partial)^2}\frac{\bar{n}^\nu F_{\nu j}}{(\bar{n}\partial)}\right]D_i\psi\right)+\frac{ic_{10}}{m_e^2}\left(D_i{\psi^\dagger}\left[\frac{n^\mu F_{i\mu}}{(\bar{n}\partial)}\frac{\bar{n}^\nu n^\alpha F_{\nu\alpha}}{(\bar{n}\partial)}+\frac{\bar{n}^\mu n^\nu F_{\mu\nu}}{(\bar{n}\partial)}\frac{n^\alpha F_{i\alpha}}{(\bar{n}\partial)}\right]\psi\right. \nonumber \\
&-&\left.\psi^\dagger\left[\frac{n^\mu F_{i\mu}}{(\bar{n}\partial)}\frac{\bar{n}^\nu n^\alpha F_{\nu\alpha}}{(\bar{n}\partial)}+\frac{\bar{n}^\mu n^\nu F_{\mu\nu}}{(\bar{n}\partial)}\frac{n^\alpha F_{i\alpha}}{(\bar{n}\partial)}\right]D_i\psi\right) \nonumber \\
&+&\frac{1}{m_e^3}\left(D^2_{jj}\psi^\dagger\left[c_{11}\frac{\bar{n}^\mu F_{\mu i}}{(\bar{n}\partial)}\frac{\bar{n}^\nu F_{\nu i}}{(\bar{n}\partial)}+c_{12}\frac{\bar{n}^\mu n^\nu F_{\mu\nu}}{(\bar{n}\partial)}\frac{\bar{n}^\alpha n^\beta F_{\alpha\beta}}{(\bar{n}\partial)}\right]\psi\right. \nonumber \\
&+&\left.\psi^\dagger\left[c_{11}\frac{\bar{n}^\mu F_{\mu i}}{(\bar{n}\partial)}\frac{\bar{n}^\nu F_{\nu i}}{(\bar{n}\partial)}+c_{12}\frac{\bar{n}^\mu n^\nu F_{\mu\nu}}{(\bar{n}\partial)}\frac{\bar{n}^\alpha n^\beta F_{\alpha\beta}}{(\bar{n}\partial)}\right]D^2_{jj}\psi\right)\nonumber\\
&+&\frac{1}{m_e^3}D_j\psi^\dagger\left[c_{13}\frac{\bar{n}^\mu F_{\mu i}}{(\bar{n}\partial)}\frac{\bar{n}^\nu F_{\nu i}}{(\bar{n}\partial)}+c_{14}\frac{\bar{n}^\mu n^\nu F_{\mu\nu}}{(\bar{n}\partial)}\frac{\bar{n}^\alpha n^\beta F_{\alpha\beta}}{(\bar{n}\partial)}\right]D_j\psi \nonumber \\
&+&\frac{c_{15}}{m_e^3}\left(D^2_{33}\psi^\dagger\frac{\bar{n}^\mu n^\nu F_{\mu\nu}}{(\bar{n}\partial)}\frac{\bar{n}^\alpha n^\beta F_{\alpha\beta}}{(\bar{n}\partial)}\psi+\psi^\dagger\frac{\bar{n}^\mu n^\nu F_{\mu\nu}}{(\bar{n}\partial)}\frac{\bar{n}^\alpha n^\beta F_{\alpha\beta}}{(\bar{n}\partial)}D^2_{33}\psi\right) \nonumber \\
&+&\frac{c_{16}}{m_e^3}D_3\psi^\dagger\frac{\bar{n}^\mu n^\nu F_{\mu\nu}}{(\bar{n}\partial)}\frac{\bar{n}^\alpha n^\beta F_{\alpha\beta}}{(\bar{n}\partial)}D_3\psi+\frac{c_{17}}{m_e^3}\left(D^2_{3i}\psi^\dagger\left[\frac{\bar{n}^\mu n^\nu F_{\mu\nu}}{(\bar{n}\partial)}\frac{\bar{n}^\alpha F_{\alpha i}}{(\bar{n}\partial)}+\frac{\bar{n}^\mu F_{\mu i}}{(\bar{n}\partial)}\frac{\bar{n}^\nu n^\alpha F_{\nu\alpha}}{(\bar{n}\partial)}\right]\psi\right. \nonumber \\
&-&\left.D_{3}\psi^\dagger\left[\frac{\bar{n}^\mu n^\nu F_{\mu\nu}}{(\bar{n}\partial)}\frac{\bar{n}^\alpha F_{\alpha i}}{(\bar{n}\partial)}+\frac{\bar{n}^\mu F_{\mu i}}{(\bar{n}\partial)}\frac{\bar{n}^\nu n^\alpha F_{\nu\alpha}}{(\bar{n}\partial)}\right]D_i\psi-D_i\psi^\dagger\left[\frac{\bar{n}^\mu n^\nu F_{\mu\nu}}{(\bar{n}\partial)}\frac{\bar{n}^\alpha F_{\alpha i}}{(\bar{n}\partial)}+\frac{\bar{n}^\mu F_{\mu i}}{(\bar{n}\partial)}\frac{\bar{n}^\nu n^\alpha F_{\nu\alpha}}{(\bar{n}\partial)}\right]D_3\psi\right. \nonumber \\
&+&\left.\psi^\dagger\left[\frac{\bar{n}^\mu n^\nu F_{\mu\nu}}{(\bar{n}\partial)}\frac{\bar{n}^\alpha F_{\alpha i}}{(\bar{n}\partial)}+\frac{\bar{n}^\mu F_{\mu i}}{(\bar{n}\partial)}\frac{\bar{n}^\nu n^\alpha F_{\nu\alpha}}{(\bar{n}\partial)}\right]D^2_{3i}\psi\right)\,.
\end{eqnarray}}
The power counting for this expression is the same as in Eq.~(\ref{lagrascet1}). However,
we are considering now higher velocities, and hence the relative size of $\frac{1-v}{1+v}$ and $\alpha$ differs from the 
previous case. For $\frac{1-v}{1+v}\gg\alpha^2\gg\left(\frac{1-v}{1+v}\right)^{3/2}$ we have listed above all the operators
up to the order of $m_e\alpha^5$.
In practice, only the operators proportional to $c_1$, $c_2$, $c_{11}$ and $c_{13}$ contribute to our calculation.
The Wilson coefficients 
are fixed by the matching calculation sketched in~(\ref{match2}). We obtain
\begin{eqnarray}
c_1&=&c_2=c_6=c_9=\frac{e^2}{2}\,, \\
c_3&=&c_4=c_{10}=c_8=-\frac{e^2}{2}\,, \\
c_5&=&c_7=c_{16}=\frac{e^2}{4}\,, \\
c_{11}&=&c_{12}=\frac{3e^2}{16}\,, \\
c_{13}&=&c_{14}=c_{17}=\frac{e^2}{8}\,, \\
c_{15}&=&-\frac{3e^2}{8}\,.
\end{eqnarray}

\end{document}